\pdfoutput=1
\documentclass[10pt,twocolumn,preprintnumbers,amsmath,amssymb,nofootinbib
,superscriptaddress]{revtex4-1}
\usepackage{graphicx,longtable,mathrsfs,color,array}
\usepackage[hidelinks]{hyperref}
\usepackage[usenames,dvipsnames]{xcolor} 
\usepackage{amssymb,amsmath,mathtools,mathrsfs,slashed} 
\usepackage{epsfig,subfigure,placeins,float} 
\usepackage{booktabs,longtable,ctable,multirow} 
\usepackage{exscale,relsize} 
\usepackage[normalem]{ulem} 
\usepackage{enumerate}
\usepackage{times, mathptmx} 
\usepackage[utf8]{inputenc}
\usepackage{color}
\usepackage{graphicx}
\usepackage{color}
\usepackage{graphicx,graphics}
\usepackage{bm}
\usepackage{tabularx}

\usepackage{natbib}
\setcitestyle{authoryear}

\usepackage{units}
\allowdisplaybreaks[1]
\usepackage{hhline}

\begin{document}
\title{The dust mass in Cassiopeia A from infrared and optical line flux differences}


\author{Maria Niculescu-Duvaz}
\affiliation{Dept. of Physics \& Astronomy, University College London, Gower St, London WC1E 6BT,UK}
\author{Michael J. Barlow}
\affiliation{Dept. of Physics \& Astronomy, University College London, Gower St, London WC1E 6BT,UK}
\author{Antonia Bevan}
\affiliation{Dept. of Physics \& Astronomy, University College London, Gower St, London WC1E 6BT,UK}
\author{Danny Milisavljevic}{
\affiliation{Department of Physics and Astronomy, Purdue University, 525 Northwestern Ave., West Lafayette, IN 47907, USA}
\author{Ilse De Looze}
\affiliation{Sterrenkundig Observatorium, Ghent University, Krijgslaan 281 - S9, 9000 Gent, Belgium}

\begin{abstract}
The large quantities of dust that have been found in a number of high redshift galaxies have led to suggestions that core-collapse supernovae (CCSNe) are the main sources of their dust and have motivated the measurement of the dust masses formed by local CCSNe. For Cassiopeia~A, an oxygen-rich remnant of a Type~IIb CCSN, a dust mass of 0.6-1.1~M$_\odot$ has already been determined by two different methods, namely (a) from its far-infrared spectral energy distribution and (b) from analysis of the red-blue emission line asymmetries in its integrated optical spectrum. We present a third, independent, method for determining the mass of dust contained within Cas~A. This compares the relative fluxes measured in similar apertures from [O~{\sc iii}] far-infrared and visual-region emission lines, taking into account foreground dust extinction, in order to determine internal dust optical depths, from which corresponding dust masses can be obtained. Using this method we determine a dust mass within Cas~A of at least 0.99$^{+0.10}_{-0.09}$~M$_\odot$.

\end{abstract}
\maketitle



\section{Introduction}

The discovery of large quantities of dust in a number of high-redshift (z $>$ 6) galaxies and quasars \citep[e.g.][]{Bertoldi2003, Watson2015, Laporte2017} prompted a shift away from AGB stars being perceived as the primary dust factories in the Universe.
Instead, it has been proposed that a significant fraction of cosmic dust, particularly at high redshifts, is formed in the ejecta of core-collapse supernovae (CCSNe), with  \cite{Morgan2003} and \cite{Dwek2007} estimating that each CCSN would need to produce $\geq$0.1 M$_\odot$ of dust for this to be the case. \\

Typically, the dust mass in CCSNe and supernova remnants (SNRs) has been estimated by fitting the dust spectral energy distribution (SED) at infrared wavelengths. {\em Kuiper Airborne Observatory} and {\em Spitzer Space Telescope} mid-infrared observations of CCSNe ejecta made up to three years after outburst typically found warm dust masses of only $10^{-4}$ - $10^{-3}$ M$_\odot$ to be present \citep[e.g.][]{Wooden1993, Sugerman2006, Kotak2009, Fabbri2011}. However, \cite{Matsuura2011} utilised {\em Herschel Space Observatory} observations of SN~1987A taken 23 years after outburst to probe previously undetectable T$\sim$23~K cold dust emitting at far-IR wavelengths and derived a cold dust mass of $\sim$0.5~M$_\odot$. Follow-up ALMA observations of SN~1987A \citep{Indebetouw2014} resolved this dust component to be at the centre of the remnant. 

The 340-year old oxygen-rich supernova remnant Cas~A has had a series of infrared-based measurements made of its dust mass and dust composition, with the derived mass increasing as observations at progressively longer wavelengths exposed emission from increasingly cooler dust. A broad 21 $\mu$m emission feature in the {\em ISO}-SWS spectrum of Cas~A was identified as a silicate species by \cite{Arendt_1999}. This feature and its correlations with a range of Cas~A physical properties has been studied in detail by a number of subsequent papers \citep{douvion_2001, Ennis_2006, Rho2008, Arendt_2014}.
From {\em ISO} observations obtained out to 30 $\mu$m, \cite{douvion_2001} derived a mass of $\sim$90~K warm dust of at least 10$^{-4}$~M$_\odot$. 
From a fit to {\em IRAS} 60- and 100 $\mu$m fluxes, \cite{Arendt_1999} estimated 0.038~M$_\odot$ of 52~K dust to be present in Cas~A.
From far-infrared and submillimetre photometry out to 500~$\mu$m, a $\sim$35~K cold dust mass of 0.06~M$_{\odot}$ was estimated for Cas~A
by \cite{Sibthorpe_2010} using {\em AKARI} and {\em BLAST} photometry, while 0.075~M$_\odot$ of $\sim$35~K dust was derived for Cas~A by \cite{Barlow2010} using {\em Herschel} PACS and SPIRE photometry over a similar wavelength range.  \cite{Arendt_2014} analysed {\em Spitzer} and {\em Herschel}-PACS data out to 160~$\mu$m and found $\leq$0.1~M$_{\odot}$ of dust emitting out to that wavelength.
\cite{DeLooze2017} (DL2017) conducted a spatially resolved study of the dust in Cassiopeia~A (Cas~A) using {\em Spitzer} and {\em Herschel} data out to 500 $\mu$m. They found that the largest proportion of the dust mass was in a cold dust component, which, similar to SN~1987A, resided in the unshocked interior of Cas~A. They derived the total ejecta dust mass in Cas~A to be 0.5$\pm$0.1~M$_\odot$. \cite{Priestley2019}, modelled the thermal emission dust grains in Cas~A heated by synchrotron radiation and particle collisions and found a dust mass in agreement with the DL2017 estimate. 
These results hint at the possibility that CCSNe could be the main contributors to the total dust budget of the Universe. A larger sample of dust masses for a diverse range of CCSNe and SNRs, and measured with a diverse range of techniques, together with tighter constraints on dust destruction rates in SNR reverse shocks \citep[e.g.][]{Kirchschlager2019, Kirchschlager2020, slav2020} is needed to confirm this view.

As {\em Herschel} is no longer functioning, different methods now have to be used if we are to determine dust masses for a larger sample of CCSNe and SNRs. \cite{Lucy1989} showed that one can determine the mass of dust that has condensed in CCSNe ejecta by exploiting the red-blue asymmetries in the optical broad line profiles. This effect is created by light from the receding red-shifted side being absorbed by more dust than light from the approaching blue-shifted side. A.~Bevan
developed the Monte Carlo radiative transfer code {\sc damocles} \citep{Bevan2016}, which models the red-blue asymmetries in CCSNe and SNRs to determine their dust properties. The dust masses deduced using it for SN~1987A \citep{Bevan2016} and Cas~A \citep{Bevan2017} are in good agreement with the dust masses from {\em Herschel}-based analyses of the far-IR dust emission from SN~1987A and Cas~A \citep[][respectively]{Matsuura2011, DeLooze2017}. 

Cas~A is a young, oxygen-rich supernova remnant (SNR) with an age of roughly 340 years \citep{Fesen2006}. 
An expansion distance of 3.33$\pm$0.10~kpc has been determined by \cite{Alarie2014}, in good agreement with the earlier expansion distance measurement by
\cite{Reed1995} of 3.4$^{+0.3}_{-0.1}$~kpc. 
It is the result of a Type IIb supernova,  classified by \cite{Krause2008} from a spectral identification of light from the original explosion echoing off interstellar material. Progenitor mass estimates for Cas~A range from 15 - 25 M$_\odot$  \citep[e.g.][]{Young2006}.
It lies in the Perseus spiral arm of the Milky Way and the light we receive from it experiences a large amount of interstellar extinction. \cite{Dunne2003} attributed a cold dust component of around 3 M$_\odot$ to Cas A from SCUBA submillimetre polarimetric observations, but it has been argued that most of the sub-mm emission was actually from cold dust in a foreground molecular cloud complex \citep{Krause2004, Wilson2005}. Whether there is an interaction between Cas A and nearby molecular clouds is also disputed. \cite{Ma2019} and \cite{Kilpatrick2014} detected some CO-emitting regions around Cas~A which they believed could indicate an interaction, whereas \cite{Zhou_2018} argued that the molecular cloud complex is in the foreground of Cas~A and not interacting with it. 

From X-ray observations, the ejecta has been observed to be impacted by a reverse shock \citep{McKee1974}.
The kinematic 3D structure of Cas A's ejecta has been studied by several authors. The optically emitting ejecta has been mapped by \cite{Reed1995}, by \cite{Milisavljevic2013} (M2013) and by \cite{Alarie2014}.
The 3D doppler reconstruction of M2013 is comprised of ~1 pc diameter rings, which they interpreted as cross sections of bubbles created by radioactive $^{56}$Ni-rich ejecta inflating and compressing material. The \cite{Milisavljevic2015} study of the interior unshocked ejecta emitting in the near-IR  supports this interpretation.
\cite{DeLaney2010} also created a 3D doppler reconstruction of Cas A using {\em Spitzer} data, mapping the reverse-shocked ring emitting in [Ar~{\sc ii}] and [Ne~{\sc iii}], as well as the unshocked interior emission which is bright in  [Si~{\sc ii}]. This led them to adopt a disk model for Cas~A tilted away from the plane of the sky. Assuming the reverse shock is spherical, they posited that the fact that no optical emission is seen in the centre of Cas A is because the reverse shock is currently interacting with the edges of the tilted disk, and has not reached the centre yet.

\begin{figure*}
	\includegraphics[width=\textwidth]{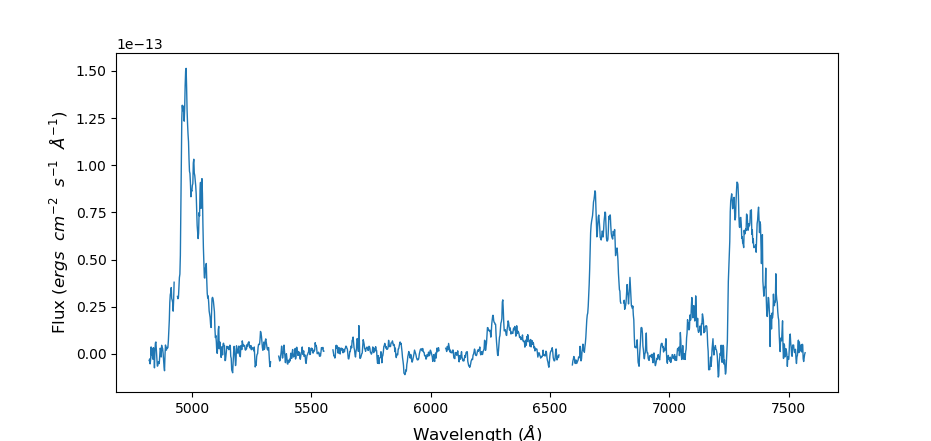}
    \caption{The integrated spectrum of Cas~A from the M2013 dataset, before correcting for interstellar extinction. For each slit placement, 450 one-dimensional spectra in apertures extracted from the long-slit spectra were co-added, then all 90 slit spectra were added together. The main emission features are the [O~{\sc iii}] 4959,5007 \AA\ doublet, the [O~{\sc i}] 6300,6363 \AA\ doublet, the [S~{\sc ii}] 6717,6731 \AA\ doublet, [Ar~{\sc iii}] 7136 \AA\ and the [O~{\sc ii}] 7320,7331 \AA\ doublet. Gaps in the spectra are from where contaminating sky lines have been removed.}
    \label{fig:integ_casa}
\end{figure*}

The goal of the current work is to use a new method to
determine ejecta dust masses at multiple locations around Cas~A, by comparing the relative intensities of far-infrared and optical forbidden emission lines in the same apertures from the same ions of oxygen, either neutral or doubly ionized. We assume that the infrared line intensities are unaffected by dust extinction and reflect the intrinsic distributions of the given species in Cas A, and that any deviations in the intensities of the optical forbidden lines (after correction for significant interstellar extinction) are due to internal dust extinction.

In Section 2 below we describe our optical and infrared line measurements. In Section 3 we describe how we correct our line fluxes for interstellar extinction. We then use the infrared fine structure line fluxes to predict the intrinsic optical line fluxes from the same ions, and compare the predicted optical line fluxes with the interstellar extinction-corrected line fluxes to determine internal dust optical depths and dust masses. In Section 4 we discuss a number of statistical and systematic sources of uncertainty for our derived quantities.
In Section 5 we discuss the future survivability of the dust in Cas~A while in Section 6 we present our conclusions.

\section{Observational Data}

\subsection{MDM Modular Spectrograph optical spectra}
Two of the datasets used to make the 3D kinematic reconstruction of Cas~A by M2013 were used in this work. These were taken in September 2007 and September 2008 and consisted of 58 and 45 long slit optical spectra respectively, covering the entire optically bright ring.  The spectra were taken with the MDM Modular spectrograph on the Hiltner 2.4m telescope at Kitt Peak. The dimensions of the long slit were ${2^{\prime \prime} \times 5^{\prime}}$, and   
successive slit positions were spaced by $3^{\prime \prime}$ , orientated from North to South. The spectra covered a wavelength range of  4500-7700 \AA\  with a resolution of 6 \AA , corresponding to a velocity resolution of 360~km~s$^{-1}$ at the air wavelength of the 5006.843 \AA\ transition of [O~{\sc iii}]. The exposure times for each long slit spectrum were 1000s. The slit placements are shown in Figure 1 of M2013. An integrated spectrum of Cas A obtained by co-adding the 2007 and 2008 long-slit datasets is shown in Figure \ref{fig:integ_casa}.

J2000 co-ordinates were assigned to each position. To illustrate the optical structure of Cas A, ejecta knots were identified by the presence of broad [O~{\sc iii}] emission, as shown by the grey and coloured dots in Figures  ~\ref{fig:2d_casa_iso_op} and ~\ref{fig:2d_casa_pacs_op}, where emission knots falling within each of the ISO-LWS and PACS apertures are coloured.
Ballistic trajectories from the centre of expansion \citep{Thorstensen2001} were assumed for all the ejecta knots.
M2013 fitted a spherical expansion model to all of the identified ejecta knots, deriving a scale factor to convert from angular distance from the center of expansion (COE) to transverse velocity of 0.022~arcsec per km~s$^{-1}$.

\subsection{{\em ISO}-LWS Far-Infrared Spectra}

Spectra of Cas~A were taken in 1997 using the {\em Infrared Space Observatory} Long Wavelength Spectrometer 
\citep[{\em ISO} LWS;][]{Clegg1996, Swinyard1996} with six pointings on Cas~A and one on offset position 4, using an 84$''$ diameter circular aperture.  The spectra had a spectral resolution of 0.3 $\mu$m\ from 43-92 $\mu$m , 
corresponding to velocity resolutions of 1737, 1424 and 1020~km~s$^{-1}$ at the vacuum wavelengths of the 51.815, 63.185 and 88.356 $\mu$m lines, respectively. The spectra were retrieved from the {\em ISO} LWS archive, in a data type called ``uniformly processed LWS L01 spectra''. The observations are summarized in Table 2 of \cite{Docenko2010}. The positions of the six LWS apertures are shown in Figure~\ref{fig:2d_casa_iso_op} in a transverse velocity frame, where the velocities of the centre of the aperture are given with respect to the same centre of expansion (COE) as for the optical data.

\subsection{PACS-IFU Far-Infrared Spectra}

Spectroscopic observations with the {\em Herschel} PACS-IFU \citep{Poglitsch2010} were taken in chopping mode for 9 regions around Cas~A, on January 1st 2011. The central WCS co-ordinate of each aperture is summarized in Table 1 of DL2017. The PACS-IFU apertures were orientated at a position angle of 240$^{\circ}$ and are shown in Figure \ref{fig:2d_casa_pacs_op}. We made use of spectra in the wavelength range 70-105  $\mu$m, taken with the SED Range Mode B2B + Long R1. The spectral resolution at 90 $\mu$m was 120~km~s$^{-1}$.
The PACS spectra had been reduced using the standard PACS chopped large range scan pipeline in HIPE v14.0.0, with the PACS CAL 32 0 calibration file. The flux calibration uncertainty for the PACS line measurements was assumed to be 13~per~cent for our wavelength region of interest. 

\subsection{Spectral Processing}
The optical and IR spectra were manually continuum-corrected using the {\sc dipso} package \citep{Howarth2004}. The twenty-five $9.4'' \times 9.4''$ spaxels in each PACS-IFU pointing were co-added. 

All spaxels from the M2013 optical dataset that spatially coincided with each {\em ISO}-LWS and PACS-IFU aperture were summed together. The optical spectra were convolved to have the same spectral resolution of the {\em ISO}-LWS spectrum of the 52 $\mu$m line, which was 1700~km~s$^{-1}$.
The [O~{\sc iii}] 4959,5007 \AA and [O~{\sc i}] 6300,6363 \AA doublets were extracted from the optical spectra. Adopting an intrinsic [O~{\sc i}]6300/6363 \AA\ intensity ratio of 3.13 \citep{Baluja_1988}, the [O~{\sc i}] 6363 \AA\ component was deblended from the [O~{\sc i}] 6300 \AA\ component. The 4959 \AA\ contribution to the [O~{\sc iii}] 4959,5007 \AA\ doublet was also removed, assuming a doublet intensity ratio of 1:2.98 \citep{Storey2000}. 
All optical and infrared integrated fluxes of the continuum-corrected {\em ISO}-LWS and PACS-IFU spectra were calculated using the FLUX utility in {\sc dipso}.

\begin{figure}

	\includegraphics[width=\columnwidth]{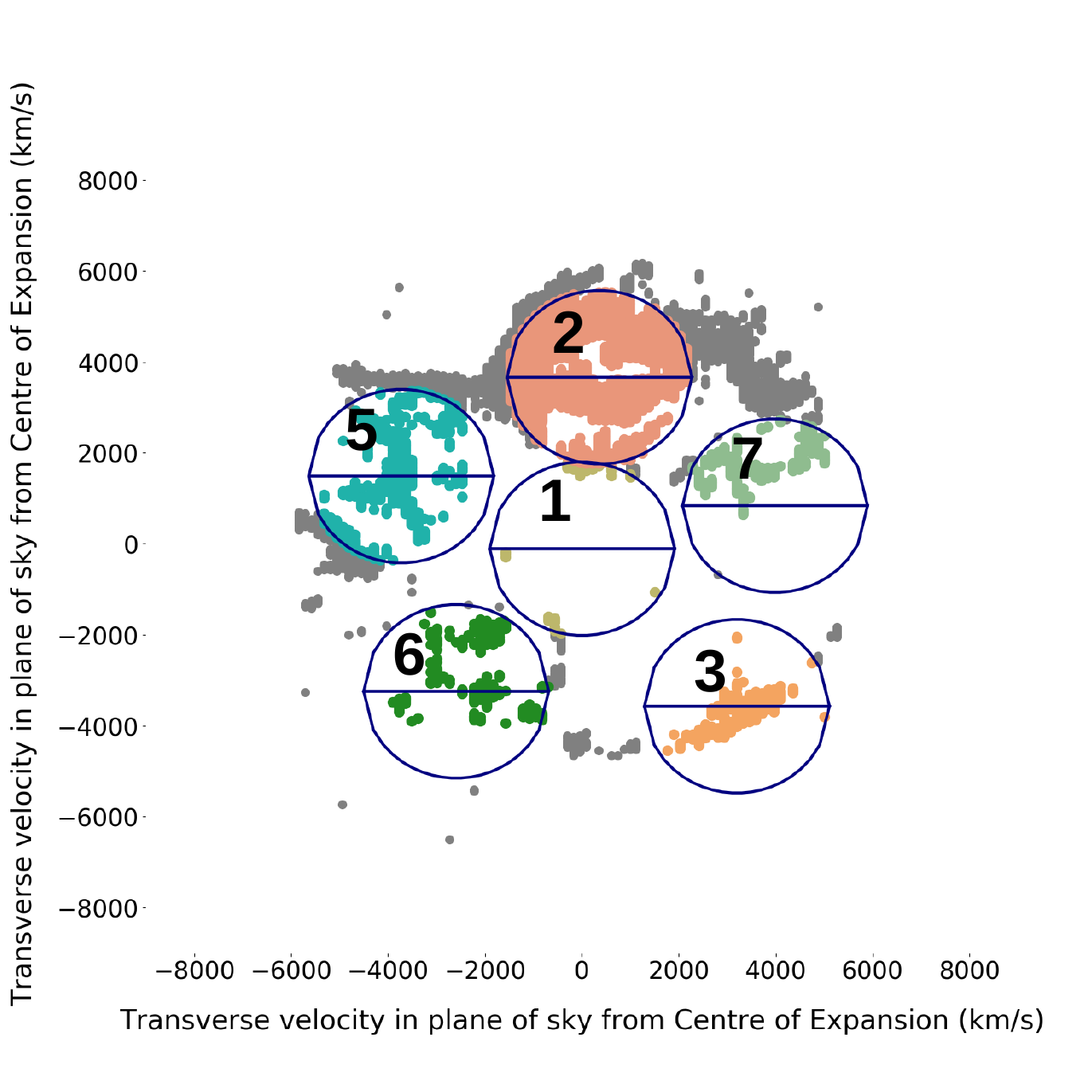}
    \caption{The circles show the {\em ISO}-LWS pointing apertures, numbered by how they appear in the archive. The WCS centre co-ordinates were determined from the {\em ISO} LWS archive, and converted to velocities from the centre of expansion using the same M2013 technique as for the optical dataset. Filled points show where [O~{\sc iii}] 5007 \AA\ emission was detected in spectra from the M2013 dataset. The multi-coloured points show where such emission fell within each of the {\em ISO}-LWS apertures.}
    \label{fig:2d_casa_iso_op}
\end{figure}

\begin{figure}
	\hspace*{-0.65cm}
	\includegraphics[width=4in]{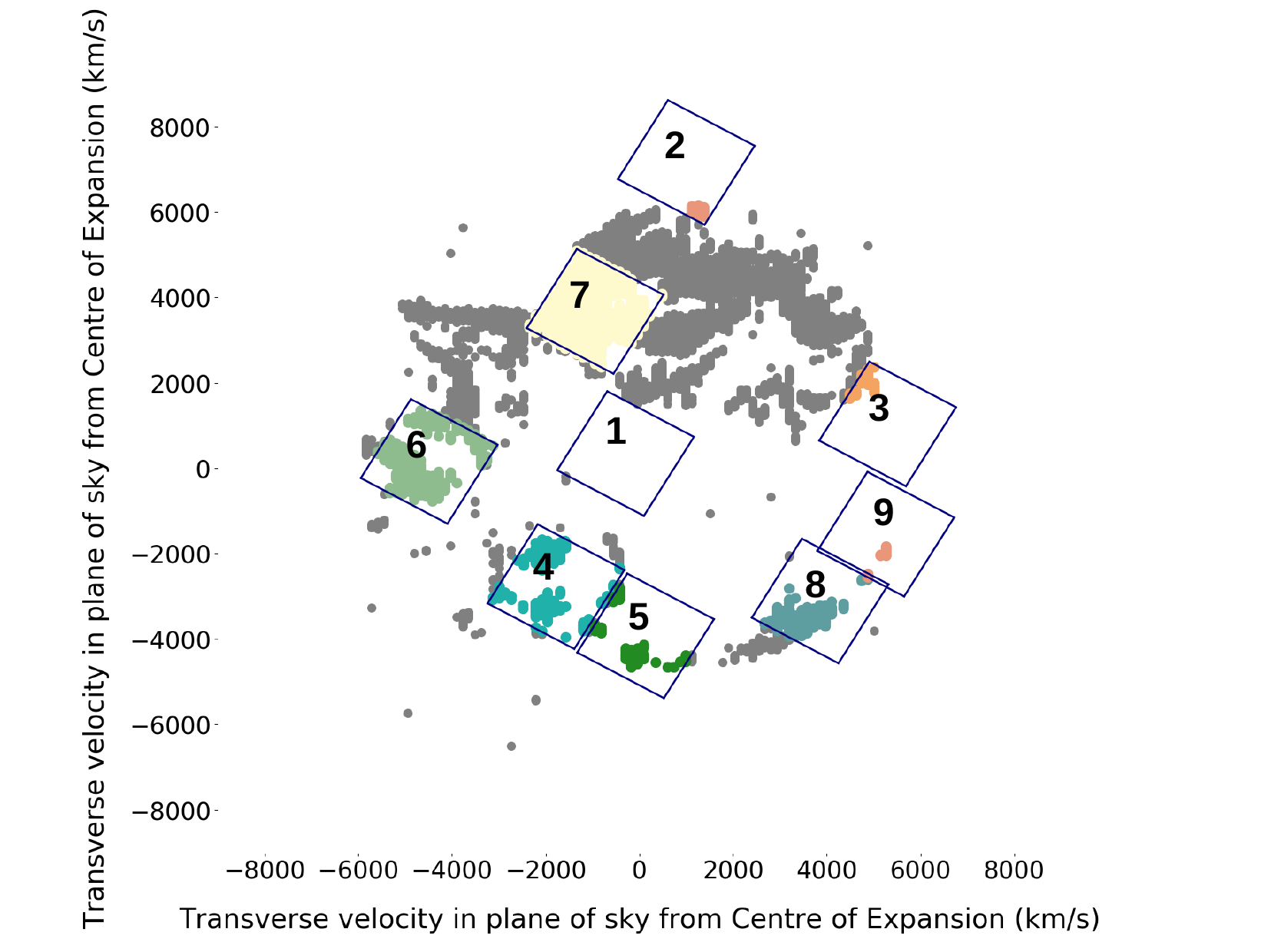}
	\hspace*{0.65cm}
    \caption{The diamonds show the PACS-IFU pointing apertures for Cas A. The WCS centre co-ordinates were taken from Table 1 of DL2017 and converted to velocities from the centre of expansion using the same M2013 technique as for the optical dataset. Filled points show where [O~{\sc iii}] 5007 \AA\ emission was detected in the M2013 dataset. The multi-coloured points show where such emission fell within each of the PACS apertures.}
    \label{fig:2d_casa_pacs_op}
\end{figure}

\subsection{{\em ISO}-LWS and PACS-IFU line profile comparison}

Aperture~6 of the {\em ISO} LWS dataset and aperture 4 of the PACS-IFU dataset cover very similar regions of Cas~A, as seen in Figures~\ref{fig:2d_casa_iso_op} and ~\ref{fig:2d_casa_pacs_op}.
Figure~\ref{fig:iso_ap6_pacs_ap4} shows the [O~{\sc iii}] 88 $\mu$m line profiles from these overlapping apertures of the LWS and PACS datasets. The spectrum from PACS aperture 4 has been convolved to the resolution of the {\em ISO} spectra. Both the line shapes and their integrated fluxes agree quite well, considering the different aperture sizes. The flux of the [O~{\sc iii}] 88 $\mu$m line in LWS aperture 6 was 4.34 $\times$ 10$^{-11}$ ergs~$cm^{-2} s^{-1}$  and in PACS-IFU aperture 4 was 2.81 $\times$ 10$^{-11}$ ergs~$cm^{-2} s^{-1}$.

\begin{figure}
	\includegraphics[width=\columnwidth]{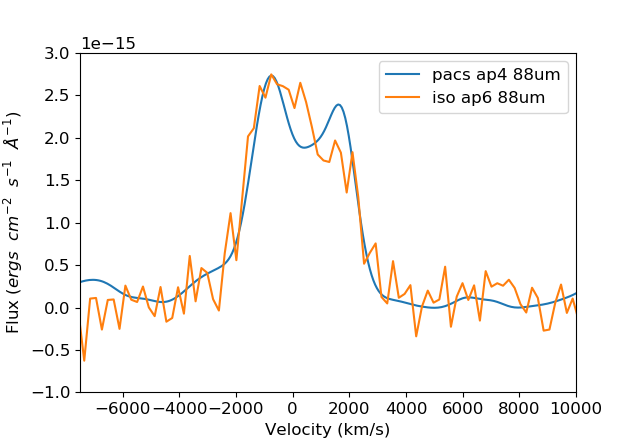}
    \caption{The [O~{\sc iii}] 88 $\mu$m line profile in {\em Herschel} PACS-IFU aperture 4 and in {\em ISO}-LWS aperture 6. The PACS spectrum has been convolved to the resolution of the LWS spectrum.}
    \label{fig:iso_ap6_pacs_ap4}
\end{figure}

\begin{figure}
	\includegraphics[angle=-90,width=0.49\textwidth]{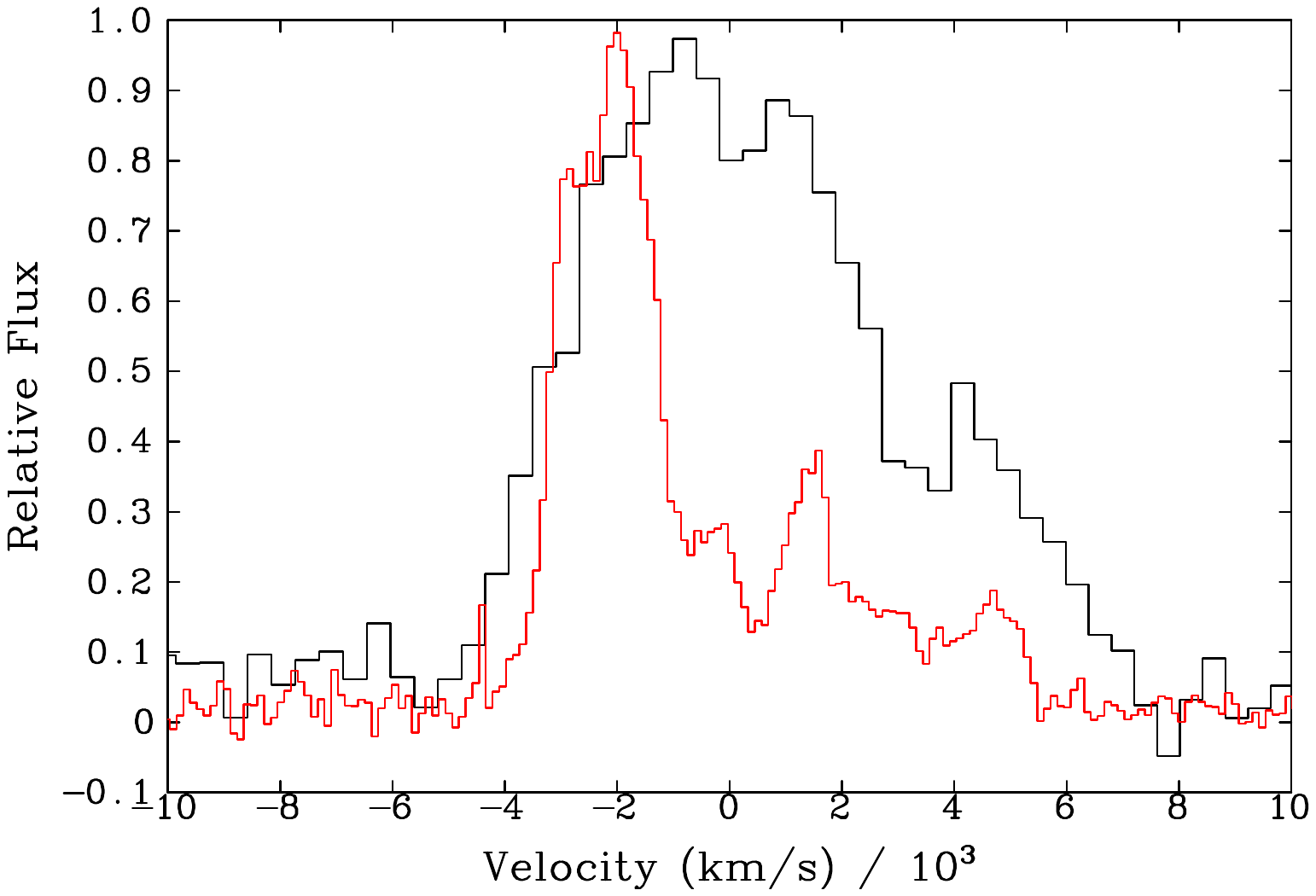}
    \caption{
    The [O~{\sc iii}] 52 $\mu$m line profile (black) summed over all six {\em ISO}-LWS apertures, compared to the 5007 \AA\ line profile (red) corresponding to a summation over the same projected aperture areas. The two profiles have been normalised to the same peak value. It can be seen that relative to the 52 $\mu$m profile the optical 5007 \AA\ profile has a deficit of redshifted emission, attributable to internal dust extinction.}
    \label{fig:all6_52um_5007A}
\end{figure}

\section{Determining dust masses from optical and IR line flux differences}

The procedure we will follow is based on using the measured IR fine structure line fluxes of [O~{\sc iii}] and [O~{\sc i}] to predict unreddened optical line fluxes from the same species and then comparing these predicted fluxes with those observed through the same aperture areas in order to estimate the optical extinction affecting each aperture. This extinction is assumed to consist of two components - the first being foreground interstellar dust extinction and the second being extinction due to dust internal to the remnant. In the next subsection we describe how we estimate the foreground interstellar extinction for each aperture position. Our internal dust extinction
estimate implicitly places the internal dust in a screen near the front of the remnant rather than being mixed with the emitting gas throughout the remnant. However we show in Section~4.3, via numerical modelling of gas and dust mixed in either smooth or clumped distributions, that the dust optical depths measured by our empirical method should be within 20~percent of actual dust optical depths when the dust is mixed with the emitting gas throughout the remnant. One prediction from internal dust extinction models is that redshifted gaseous line emission from the receding farside of the remnant should undergo more extinction by internal dust when traversing the nebula to the observer, than will blueshifted line emission from the approaching nearside of
the remnant. This prediction is confirmed by Figure~\ref{fig:all6_52um_5007A}, which shows the [O~{\sc iii}] 52 $\mu$m velocity profile summed over the six LWS apertures and compares it with the 5007 \AA\ velocity profile summed over the same six areas. The 5007 \AA\ line is clearly missing flux on the reshifted side of its profile. \\
\indent Cas~A has a complex internal structure \citep[see e.g.][]{Fesen2001} but our method makes no assumptions about where the doubly ionized and neutral oxygen species are located along the lines of sight projected through the remnant by the observing apertures, e.g. whether they are centrally or uniformly distributed, or whether in clumped or unclumped regions. The measured emission line fluxes are summations along the lines of sight of the emission from the various regions encountered, weighted by their electron densities and (in the case of the optical lines) by their electron temperatures. The values derived for these two quantities reflect these weightings, with electron densities derived from [O~{\sc iii}] 52/88 $\mu$m flux ratios being insensitive to the electron temperature T$_e$, while the 52 $\mu$m/5007 \AA\ T$_e$-diagnostic ratio will be weighted to lower T$_e$ values than the classic [O~{\sc iii}] 5007/4363 \AA\ optical-only line diagnostic ratio.

\cite{Priestley2019} modelled the heating sources for Cas~A's dust emission and fitted the overall IR SED with four dust components.
The two warmest dust components, accounting for less than 1~percent of the total dust mass between them, were respectively heated by the X-ray emitting diffuse reverse shocked ejecta and by the X-ray emitting material swept up by the outer blast wave.
A cold `pre-reverse shock' dust component heated by synchrotron radiation accounted for 90~percent of the dust mass. Their `clumped' gas component, with $n_e = 480$~cm$^{-3}$ and T$_e = 10^4$~K, corresponded most closely to the [O~{\sc iii}] emission analysed here, with the dust component heated by it accounting for only 10~per~cent of the total dust mass. The synchrotron-heated cold dust component is expected to account for most of the internal optical obscuration.

\subsection{Dereddening the optical line fluxes for interstellar extinction}
It is important to correct the optical fluxes in our dataset for extinction by interstellar dust. A map of interstellar extinction towards Cas~A, shown in Figure~F4 of DL2017, gives the column density of extinction, represented by A$_V$, along sightlines near to and through Cas~A. This is plotted in Figure \ref{fig:bayestar-regions}
as a colour map, along with the distribution of the optical emission knots in Cas~A, shown as black points. The white regions show where no ISM dust columns densities could be estimated due to the corresponding far-IR dust emission being too faint (DL2017). 
For each LWS and PACS-IFU aperture, we calculated a mean A$_V$ value averaging over every pixel of the DL2017 interstellar dust extinction map which fell within each aperture. These mean extinctions range from A$_V$ = 3.9 mag for LWS aperture 2 and PACS-IFU aperture 7, both at the northern part of the remnant,
to A$_V$ = 9.1 mag for LWS aperture 3 and 10.3 mag for PACS-IFU aperture 9, both at the southwestern part of the remnant.

\begin{figure*}
    \centering
    \begin{minipage}[t]{0.47\textwidth}
        \centering
        \includegraphics[width=1\textwidth]{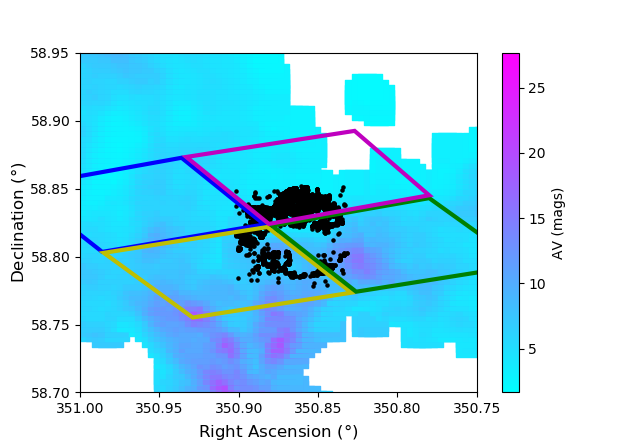} 
        \caption{The coloured rhombuses show the voxels of the G2019 3D reddening map which overlie Cas~A. They have a scale of 3.4' on a side (Table 5 in G2019). The black dots show optical emission knots in Cas~A, from the M2013 dataset. 
        The colour map shows the visual extinction of ISM dust, in magnitudes, along sightlines towards Cas~A, from Figure~F4 of DL2017.}
        \label{fig:bayestar-regions}
    \end{minipage}\hfill
    \begin{minipage}[t]{0.47\textwidth}
        \centering
        \includegraphics[width=1\textwidth]{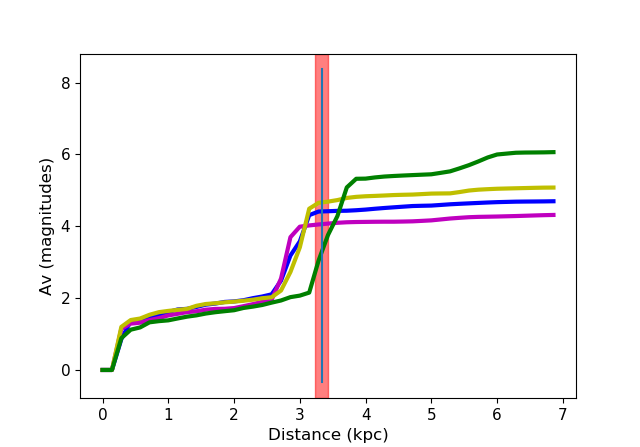} 
       \caption{Extinction in Av magnitudes as a function of distance, for the four voxels of the  G2019 map plotted as colored rhombuses in Figure \ref{fig:bayestar-regions}. The colours of the curves indicate the locations where the extinction curves are calculated, and correspond to the colours of the rhombuses in Figure~\ref{fig:bayestar-regions}.
       The vertical blue bar corresponds to the \cite{Alarie2014} distance to Cas~A of 3.33~kpc, with the broad red vertical bar corresponding to the stated errors on their estimate of this distance.}
       \label{fig:extinct-sightlines}
    \end{minipage}
\end{figure*}

Although there is no way of quantifying from this map how much of the extinction in each pixel is in front of Cas~A and how much is behind, we can show that at least a portion of the interstellar extinction, for some regions, must lie behind Cas~A. We dereddened all the optical emission knots of Cas~A with the total A$_V$ value from the closest pixel in the DL2017 ISM dust map and compared the fluxes from two [O~{\sc iii}] emission knots before and after dereddening: one in the northern region and one in the southwestern region.
Before dereddening, these had fairly similar fluxes of 9.8 $\times$ 10$^{-17}$ and 1.3 $\times$ 10$^{-16}$ ergs~$cm^{-2} s^{-1}$,  respectively. The northern knot corresponds to a total ISM A$_V$ from DL2017 of 3.6 magnitudes, while the southwestern knot corresponds to an ISM A$_V$ of 9.7 mags. 
The fluxes after dereddening were 2.7 $\times$ 10$^{-15}$ and 1.0 $\times$ 10$^{-12}$ ergs~$cm^{-2} s^{-1}$ for the northern and southwestern knot, respectively. It seems very unlikely that two optical knots, both with presumably similar density and temperature conditions, would have an intrinsic flux difference of a factor of nearly 1000, implying that at least part of some of the ISM extinction columns must lie behind Cas~A.

In an attempt to better constrain the proportions of the total interstellar dust columns that lie in front of Cas~A, we used the 3D dust reddening map created by \cite{Green2019} (G2019), to plot the reddening as a function of distance along different sightlines towards Cas~A. 
We used the BayestarWebQuery\footnote{\url{http://argonaut.skymaps.info/usage}} function in  the dustmaps python package to query a grid of 1600 galactic co-ordinates close to Cas A. The coloured rhombuses in Figure \ref{fig:bayestar-regions} show each voxel of the G2019 map around Cas A, while Figure~\ref{fig:extinct-sightlines} plots the extinction A$_{\rm V}$ versus distance relation for each voxel, where the colour of the curve matches the colour of the voxels shown in the dust reddening map. We use the distance to Cas A of \cite{Alarie2014} of $3.33^{+0.10}_{-0.10}$ kpc, and this is plotted as the blue vertical line, where the broad red bar indicates the uncertainty on this value. 

The curves in Figure~\ref{fig:extinct-sightlines} confirm the DL2017 finding that sightlines towards the western and southern regions of Cas~A have the largest interstellar dust columns. The DL2017 map also shows that the northern regions of Cas~A correspond to the lowest interstellar dust columns. From Figure~\ref{fig:extinct-sightlines} we see that for the northern region of Cas~A the relevant G2019 curve (magenta) predicts A$_{\rm V}$ = 4.0 mag at the distance of Cas~A, similar to the DL2017 prediction of a total A$_{\rm V}$ of 3.8 mag at the position of LWS aperture 2. 
For apertures falling on northern parts of the remnant (LWS aperture 2 and PACS-IFU aperture 7) we have therefore assumed that all of the interstellar dust column lies in front of the remnant.

It is notable that the other three extinction versus distance curves that are plotted in
Figure~\ref{fig:extinct-sightlines} all show a flattening at or near to the distance of Cas~A
and that they predict maximum extinctions that are significantly lower than measured for the same positions from the DL2017 map in Figure~\ref{fig:bayestar-regions}. The flattening is likely due to increasing extinction with distance beyond Cas~A leading to the G2019 algorithm running out of detectable stars with which to measure distances and extinctions.

So apart from the northern LWS-2 and PACS-7
apertures, for which we have argued that all of the DL2017 extinction column is in front of the remnant, we will assume that for the remaining apertures only 50~per~cent of the DL2017 ISM extinction columns lie in front of the remnant. This leads to adopted foreground A$_{\rm V}$ values for these apertures that are similar to those discussed above for the northernmost apertures - see column~6 of Table~\ref{table:oiii52-iso} for a comparison for A$_{5007}$.

In Section 4.2 we will consider the effect on the derived dust masses of assuming that the interstellar dust columns in front of the remnant are exactly as predicted by the G2019 extinction versus distance curves.

With the above assumptions we calculated a mean foreground A$_{\rm V}$ value for each aperture using every spaxel of the DL2017 interstellar dust extinction map which fell within a given aperture.
We used these foreground A$_{\rm V}$ values for each aperture to calculate $A_\lambda$ at 5007 \AA\ and 6300 \AA\, assuming R$_V$ = 3.1 and adopting the reddening law of \cite{Cardelli1989}. The values of $A_\lambda$ for the two optical lines and for each LWS and PACS aperture can be found in Tables ~\ref{table:oiii52-iso}, \ref{table:oiii-pacs} and \ref{table:oi-iso}. These values of $A_\lambda$ were applied to the optical fluxes in the LWS and PACS apertures, to give the ISM-dereddened F$_{5007,d}$ and F$_{6300,d}$ fluxes shown in those tables. 

\subsection{Using infrared line fluxes to estimate intrinsic optical line fluxes}

\begin{figure}
\includegraphics[width=.49\textwidth]{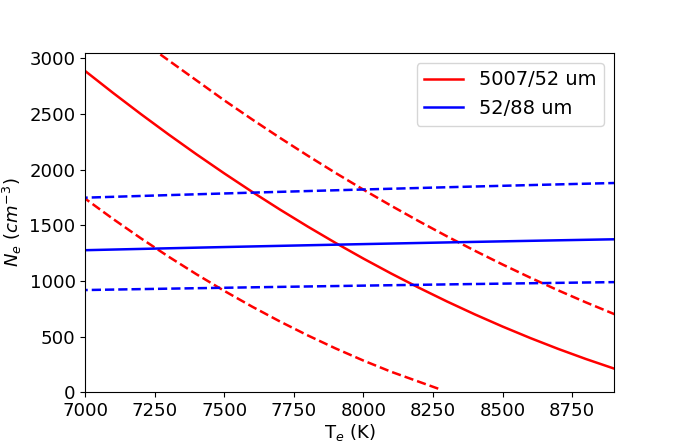}
\caption{The $n_{e}$ and T$_e$ loci corresponding to the [O~{\sc iii}] 5007 \AA /52 $\mu$m and 52/88 $\mu$m line flux ratios for LWS aperture~2 are plotted as the red and blue solid lines, respectively, with the dashed lines corresponding to the upper and lower limits on each ratio. }
\label{fig:ne-te-uncert}
\end{figure}

The 52 $\mu$m and 88 $\mu$m lines
of [O~{\sc iii}], both observed from Cas~A by the {\em ISO} LWS, are at sufficiently long wavelengths that dust extinction effects can be neglected, unlike the case of the 5007 \AA\ line of [O~{\sc iii}]. The flux ratio of the latter optical transition to either of the infrared transitions is a sensitive function of both the local electron density n$_e$ and electron temperature T$_e$ and
so we can use the IR line fluxes to predict the unreddened optical line flux if we know the values of n$_e$ and T$_e$. The 52/88 $\mu$m flux ratio is sensitive to n$_e$ across the
density range 100 $<$ n$_e$ $<$ 10$^4$ cm$^{-3}$ but is insensitive to T$_e$ due to the very low excitation energies of the two IR transitions. We have
used this flux ratio and the statistical equilibrium code {\sc equib} (written by S. Adams and I. Howarth), together with the O$^{2+}$ collision strengths of \citep{Aggarwal1983} and the transition probabilities of \citep{ Nussbaumer1981}, to derive the values of n$_e$ that correspond to each of the six LWS aperture regions
shown in Figure~\ref{fig:2d_casa_iso_op}. Column~1 of Table~\ref{table:oiii52-iso} lists the LWS aperture numbers, followed by the F$_{52}$/F$_{88}$ flux ratios and the corresponding values of n$_e$, which range from 1380~cm$^{-3}$ for LWS aperture 2 down to 115~cm$^{-3}$ for LWS aperture 1.

To predict the 5007 \AA\ line flux from an IR line flux the value of the electron temperature is also needed. A value for T$_e$ can be estimated for LWS aperture 2, for which we have argued that all of the interstellar extinction column is in front of the remnant. Our method also implicitly assumes that the internal extinction in LWS aperture 2 is very low, consistent with the low internal dust columns found in the northernmost part of the remnant by \cite[][their Fig.~F3]{DeLooze2017}
The ratio of the LWS Aperture-2 ISM-dereddened 5007 \AA\ line flux (column 7 of Table~\ref{table:oiii52-iso}) to its 52 $\mu$m line flux (column 4) yields T$_e$ = 7900$^{+400}_{-700}$~K, for n$_e$ = $1380^{+435}_{-420}$~cm$^{-3}$. Figure~\ref{fig:ne-te-uncert} shows for LWS aperture 2 the loci of the $n_{e}$ and T$_e$ solutions for the 5007 \AA /52 $\mu$m and 52/88 $\mu$m flux ratios of 2.43$^{+0.50}_{-0.68}$ and 3.25$^{+0.65}_{-0.59}$, with the dashed lines showing the loci corresponding to the upper and lower limits on the flux ratios.

Our [O~{\sc iii}] electron density and temperature derived for LWS aperture~2 are consistent with the range estimated by \cite{Docenko2010} for the same aperture.
Our [O~{\sc iii}] T$_e$ and n$_e$ values for LWS aperture 2 are also consistent with those estimated by \cite{Rho2019} from {\em SOFIA} spectra of a region in the northern part of the remnant encompassed by LWS aperture 2.
Our electron densities are consistent with the range mapped over Cas~A by \cite{Smith2009} 
using {\em Spitzer} IRS measurements of the [S~{\sc iii}] 18.7/33.6 $\mu$m line flux ratio (see their Fig.~6).

Since T$_e$ usually varies far less than n$_e$ in an ionized nebula, we have used the LWS aperture 2 value of T$_e$ = 7900~K along with the values of n$_e$ listed in column 3 to predict the 5007 \AA\ /52 $\mu$m flux ratio for each LWS aperture.
We then multiplied this ratio by the F$_{52}$ line fluxes listed in column 4 of Table~\ref{table:oiii52-iso} to obtain the expected 5007 \AA\ flux in the absence of any dust extinction, F$_{5007,exp}$, as listed in column 8. This expected flux can be compared to the ISM-dereddened flux from column 7 to obtain an estimate for the internal dust optical depth, $\tau_{int}$ (column 9), for each aperture projection through Cas~A. Table~\ref{table:oiii52-iso-uncerts} lists the statistical uncertainties for the measured and derived quantities that are listed in Table~\ref{table:oiii52-iso}; the uncertainties were combined and propagated in quadrature as appropriate.

\begin{table*}

\caption{Columns from left to right are: {\em ISO}-LWS aperture number; F$_{52}$/F$_{88}$, the observed [O~{\sc iii}] 52 $\mu$m to 88 $\mu$m flux ratio; the corresponding electron density N$_{\rm e}$ for an electron temperature of 7900~K; F$_{52}$, the observed [O~{\sc iii}] 52 $\mu$m flux; the observed [O~{\sc iii}] 5007 \AA\ flux, F$_{5007}$; A$_{5007}$, the estimated ISM extinction at 5007 \AA in magnitudes; the ISM-dereddened 5007 \AA\ flux, F$_{5007,d}$; the expected [O~{\sc iii}] 5007 \AA\ flux (F$_{5007,exp}$) in each aperture for no internal dust extinction; and the internal optical depth, $\tau_{int}$, and dust mass, M$_d$, implied by the difference between the ratios of the predicted and ISM-dereddened [O~{\sc iii}] 5007 \AA\ fluxes for cylinders projected through each {\em ISO}-LWS aperture. All fluxes are in units of ergs cm$^{-2}$s$^{-1}$ .}
\begin{tabular}{lccllllllc}
\hline
Ap& F$_{52}$/F$_{88}$ & $N_e$(cm$^{-3}$) & F$_{52}$& F$_{5007}$& A$_{5007}$ & F$_{5007,d}$ & F$_{5007,exp}$ & $\tau_{int}$ &  M$_d$ (M$_{\odot}$)\\
\hline

 2 &        3.25 &    1380 &   1.23$\times$10$^{-10}$ &   5.62$\times$10$^{-12}$ &          4.32 &   2.99$\times$10$^{-10}$ &  -  &  -  
 & -   \\ 
 1 &        0.87 &     115 &   7.42$\times$10$^{-11}$ &   8.42$\times$10$^{-14}$ &          4.85 &   7.33$\times$10$^{-12}$ &    1.23$\times$10$^{-10}$ &   2.82 &   0.31 \\ 
 3 &        2.05 &     630 &   7.47$\times$10$^{-11}$ &   2.84$\times$10$^{-13}$ &          5.11 &   3.14$\times$10$^{-11}$ &   1.41$\times$10$^{-10}$ &   1.50 &   0.16 \\ 
 5 &        1.15 &     225 &   2.77$\times$10$^{-11}$ &   5.17$\times$10$^{-13}$ &          3.48 &   1.27$\times$10$^{-11}$ &  4.62$\times$10$^{-11}$ &   1.29 &   0.14 \\ 
 6 &        2.42 &     830 &    8.30$\times$10$^{-11}$ &   4.26$\times$10$^{-13}$ &          4.21 &   2.06$\times$10$^{-11}$ &   1.69$\times$10$^{-10}$ &   2.11 &   0.23 \\ 
 7 &        1.18 &     240 &    7.60$\times$10$^{-11}$ &   5.84$\times$10$^{-13}$ &          4.32 &   3.12$\times$10$^{-11}$ &   1.27$\times$10$^{-10}$ &   1.40 &   0.15 \\ 
\hline
  & & & & & & & & Total & 0.99 \\
\hline

\label{table:oiii52-iso}
\end{tabular}
\end{table*}

\begin{table*}
\caption{Corresponding absolute uncertainties on the values in Table \ref{table:oiii52-iso}. All fluxes are in units of ergs~$cm^{-2} s^{-1}$.}
\setlength{\tabcolsep}{3pt}
\begin{tabular}{ *{19}{c} }
\hline
Ap & \multicolumn{2}{|c|}{F$_{52}$/F$_{88}$}
 & \multicolumn{2}{c|}{N$_e$}
 & \multicolumn{2}{c|}{F$_{52}$($\times$10$^{-12}$)}
& \multicolumn{2}{c|}{F$_{5007}$($\times$10$^{-14}$)}
& \multicolumn{2}{c|}{A$_{5007}$}
& \multicolumn{2}{c|}{F$_{5007,d}$($\times$10$^{-12}$)}
& \multicolumn{2}{c|}{F$_{5007,exp}$($\times$10$^{-11}$)}
& \multicolumn{2}{c|}{$\tau_{int}$}
& \multicolumn{2}{c}{M$_d$ (M$_{\odot}$)}
\\
  \hline
  & + & -  & + & - & + & - & + & - & + & - & + & - & + & - & + & - & + & -  \\
  \hline
 2 &        0.65 & 0.59 & 435 & 420 & 16 & 11 & 24 & 19 & 0.26 & 0.03 & 33.6 & 71.0 & - & - & - & - & - & -  \\ 
 1 & 0.14 & 0.13 & 55 & 50 & 4.0 & 1.7 & 3.3 & 2.3 & 0.09 & 0.23 &3.1 & 2.0 & 4.5 & 3.9 & 0.56 & 0.42 & 0.058 & 0.043\\
 3 &  0.55 & 0.41 & 290 & 195 & 8.2 & 10.6 & 5.2 & 4.7 & 0.03 & 0.00 & 5.3  & 5.0 & 1.8 & 4.8 & 0.40 & 0.38 & 0.040 & 0.037   \\ 
 5 & 0.29& 0.30 & 120 & 115 & 7.7 & 9.1 & 3.8 & 3.5 & 0.00 & 0.06 & 1.8 & 1.5 & 0.7 & 1.7 & 0.41 & 0.38 & 0.042 & 0.039\\
 6 &0.61 & 0.53 & 360 & 270 & 13.4 & 11.6 & 2.5 & 2.8 & 0.03 & 0.02 & 2.3 & 2.4 & 2.2 & 5.8 & 0.40 & 0.36 & 0.042 & 0.038\\ 
 7 &0.24 & 0.38 & 100 & 150 & 7.6 & 11.4 & 3.6 & 3.9 & 0.01 & 0.09 & 4.5 & 3.5 & 1.7 & 4.5 & 0.40 & 0.37 & 0.040 & 0.037\\
\hline
  & & & & & & & & & & & & & & & &  Total &0.10 &0.09 \\
\hline
\label{table:oiii52-iso-uncerts}
\end{tabular}
\end{table*}

\begin{table*}
\caption{Columns from left to right are: PACS aperture number; the [O~{\sc iii}] 88 $\mu$m flux; the [O~{\sc iii}] 5007 \AA\ flux; $A_{5007}$, the ISM extinction at 5007 \AA in magnitudes; the ISM-dereddened 5007 \AA\ flux; the expected optical [O~{\sc iii}] 5007 \AA\ flux in each aperture for no internal dust extinction, F$_{5007,exp}$; the  optical depth, $\tau_{int}$, and dust mass, M$_d$, implied by the ratio of the predicted and ISM-dereddened [O~{\sc iii}] 5007 \AA\ fluxes within the rectangle projected by each PACS aperture onto Cas~A. All fluxes are in units of ergs~$cm^{-2} s^{-1}$}

\begin{tabular}{lllllllc}
\hline
Ap&  F$_{88}$ & F$_{5007}$ & A$_{5007}$ & F$_{5007,d}$ & F$_{5007,exp}$& $\tau_{int}$ & M$_d$ ($M_{\odot}$) \\
 \hline
 
 1 &   3.08$\times$10$^{-11}$ &    $<$2.8$\times$10$^{-14}$ &          5.38 &   $<$4.0$\times$10$^{-12}$ &   4.46$\times$10$^{-11}$ &   $>$2.4 &   $>$0.104 \\ 
 3 &   1.10$\times$10$^{-11}$ &   1.02$\times$10$^{-13}$ &          5.13 &   1.15$\times$10$^{-11}$ &   2.19$\times$10$^{-11}$ &      0.64 & 0.028 \\ 
 4 &   2.02$\times$10$^{-11}$ &   1.82$\times$10$^{-13}$ &          3.96 &   6.98$\times$10$^{-12}$ &   1.00$\times$10$^{-10}$ &   2.67 &   0.116 \\ 
 5 &   1.32$\times$10$^{-11}$ &   4.74$\times$10$^{-14}$ &          5.13 &    5.34$\times$10$^{-12}$ &   6.55$\times$10$^{-11}$ &   2.51 &   0.109 \\ 
 6 &   4.51$\times$10$^{-12}$ &   1.92$\times$10$^{-13}$ &          4.09 &   8.30$\times$10$^{-12}$ &   9.02$\times$10$^{-12}$ &   0.08 &   - \\ 
 7 &   1.92$\times$10$^{-11}$ &   3.28$\times$10$^{-12}$ &          4.30 &   1.72$\times$10$^{-10}$ & 1.58$\times$10$^{-10}$  & -0.08 &   - \\
 
 8 &   1.27$\times$10$^{-11}$ &   1.42$\times$10$^{-13}$ &          4.96 &   1.37$\times$10$^{-11}$ &   5.04$\times$10$^{-11}$ &   1.30 &   0.056 \\ 
 9 &   8.83$\times$10$^{-12}$ &    $<$7.2$\times$10$^{-14}$ & 5.83 &  $<$1.55$\times$10$^{-11}$ &   3.42$\times$10$^{-11}$ & $>$0.79 &   $>$0.034 \\ 
 \hline
  & & & & & & Total & $>$0.45 \\
\hline
\label{table:oiii-pacs}
\end{tabular}
\end{table*}

\begin{table*}
\caption{Corresponding absolute uncertainties on the values in Table \ref{table:oiii-pacs}. All fluxes are in units of ergs~$cm^{-2} s^{-1}$.}
\begin{tabular}{ *{15}{c} }
\hline
Ap 
 & \multicolumn{2}{|c|}{F$_{88}$($\times$ $10^{-12}$) }
& \multicolumn{2}{c|}{F$_{5007}$($\times$ 10$^{-14}$)}
& \multicolumn{2}{c|}{A$_{5007}$}
& \multicolumn{2}{c|}{F$_{5007,d}$($\times$ 10$^{-12}$)}
& \multicolumn{2}{c|}{F$_{5007,exp}$($\times$ 10$^{-11}$)}
& \multicolumn{2}{c|}{$\tau_{int}$}
& \multicolumn{2}{c}{M$_d$ ($M_{\odot}$)}
\\
  \hline
  & + & - & + & - & + & - & + & - & + & - & + & - & + & -  \\
  \hline
 
 1 &  7.1 & 6.0 & 0.20 & 0.40 & 0.31 & 0.21 & 0.99 & 1.23 & 1.97 & 1.74 & 0.51 & 0.49 & 0.021 & 0.020 \\
 3 &  2.6 & 3.0 & 0.80 & 1.20 & 0.00 & 0.51 & 4.60 & 1.86 & 1.02 & 1.14 & 0.62 & 0.54 & 0.025 & 0.022\\ 
 4 &  2.4 & 2.1 & 0.90 & 1.40 & 0.03 & 0.34 & 2.72 & 1.01 & 4.37 & 3.93 & 0.59 & 0.42 & 0.024 & 0.017 \\ 
 5 & 1.9 & 1.2 & 1.02 & 1.27 & 0.19 & 0.07 & 0.88 & 1.16 & 2.91 & 2.52 & 0.47 & 0.44 & 0.019 & 0.018\\
 6 & 0.70 & 1.2 & 0.70 & 0.80 & 0.00 & 0.04 & 0.95 & 1.14 & 0.40 & 0.41 & 0.46 & 0.47 & 0.017 & 0.017\\ 
 7 &  1.6 & 2.3 & 15 & 15 & 0.00 & 0.02 & 19.1 & 23.7 & 6.1 & 5.5 & 0.40 & 0.38 & 0.017 & 0.017 \\
 8 & 1.40 & 1.94 & 2.5 & 1.9 & 0.05 & 0.10 & 2.07 & 2.41 & 2.13 & 2.10 & 0.45 & 0.45 & 0.024 & 0.025\\ 
 9 &  1.6 & 1.2 & 1.1 & 0.80 & 0.07 & 0.00 & 1.93 & 2.37 & 1.42 & 1.30 & 0.43 & 0.41 & 0.016 & 0.015\\
\hline
  & & & & & & & & & & & &  Total &0.056 & 0.052\\
\hline
\label{table:oiii-pacs-uncerts}
\end{tabular}
\end{table*}

\begin{table*}
\caption{Columns from left to right are: {\em ISO}-LWS aperture number; the observed [O~{\sc i}] 63 $\mu$m flux, F$_{63}$; the observed [O~{\sc i}] 6300 \AA\ flux, F$_{6300}$, from the same area; A$_{6300}$ the estimated ISM extinction at 6300 \AA in magnitudes; the ISM-dereddened 6300 \AA\ flux, F$_{6300,d}$; the expected optical [O~{\sc i}] 6300 \AA\ flux, F$_{6300,exp}$, in each aperture for no internal dust extinction; the internal optical depth $\tau_{int}$ and dust mass, M$_d$, implied by the difference between the predicted and ISM-dereddened [O~{\sc i}] 6300 \AA\ fluxes for cylinders projected through each {\em ISO}-LWS aperture. All fluxes are in units of ergs~$cm^{-2} s^{-1}$.}
\begin{tabular}{lllllllc}
\hline
Ap&  F$_{63}$ & F$_{6300}$ & A$_{6300}$ & F$_{6300,d}$ & F$_{6300,exp}$ & $\tau_{int}$ & M$_d$ ($M_{\odot}$)\\
\hline

 2 &   2.98$\times$10$^{-11}$ &   1.20$\times$10$^{-12}$ &          3.31 &   2.53$\times$10$^{-11}$ &   - &      - &       - \\ 
 1 &   6.35$\times$10$^{-12}$ &   2.63$\times$10$^{-14}$ &          3.72 &   8.09$\times$10$^{-13}$ &   5.40$\times$10$^{-12}$ &   1.90 &   0.31 \\ 
 3 &   1.17$\times$10$^{-11}$ &   6.61$\times$10$^{-14}$ &          3.91 &   2.42$\times$10$^{-12}$ &   9.95$\times$10$^{-12}$ &  1.41 &  0.23 \\ 
 
 5 &   1.24$\times$10$^{-12}$ &   8.85$\times$10$^{-14}$ &          2.67 &   1.04$\times$10$^{-12}$ & 1.05$\times$10$^{-12}$ & 0.01 &   - \\ 
 
 6 &   4.52$\times$10$^{-12}$ &   1.13$\times$10$^{-13}$ &          3.23 &   2.21$\times$10$^{-12}$ &   3.84$\times$10$^{-12}$ &   0.55 &   0.09 \\ 
 7 &   8.58$\times$10$^{-12}$ &   1.21$\times$10$^{-13}$ &          3.31 &      2.55$\times$10$^{-12}$ &   7.29$\times$10$^{-12}$ &   1.05 &   0.17 \\ 

 \hline
 & & & & & & Total & 0.80
\\
\hline

\label{table:oi-iso}
\end{tabular}
\end{table*}

\begin{table*}
\caption{Corresponding absolute uncertainties on the values in Table \ref{table:oi-iso}. All fluxes are in units of ergs~$cm^{-2} s^{-1}$.}
\begin{tabular}{ *{15}{c} }
\hline
Ap 
 & \multicolumn{2}{|c|}{F$_{63}(\times 10^{-12})$}
& \multicolumn{2}{c|}{F$_{6300}(\times 10^{-14})$}
& \multicolumn{2}{c|}{A$_{6300}$}
& \multicolumn{2}{c|}{F$_{6300,d}(\times 10^{-13})$}
& \multicolumn{2}{c|}{F$_{6300,exp}(\times 10^{-12})$}
& \multicolumn{2}{c|}{$\tau_{int}$}
& \multicolumn{2}{c}{M$_d$ ($M_{\odot}$)}
\\
  \hline
  & + & - & + & - & + & - & + & - & + & - & + & - & + & -  \\
  \hline
 
 2 & 1.6 & 2.2 & 5.0 & 7.7 & 0.19 & 0.02  & 27.8 & 51.1 & 4.86 & 6.70 & 0.22 & 0.33 & - & -\\
 1 & 0.64 & 0.41 & 0.74 & 0.70 & 0.07 & 0.18 & 2.44 & 1.91 & 1.09 & 1.41 & 0.36 & 0.35 & 0.059 & 0.057 \\ 
 3 &  0.80 & 0.90 & 1.1 & 1.1 & 0.02 & 0.00 & 4.20 & 4.20 & 1.94 & 2.63 & 0.26 & 0.32 & 0.042 & 0.051\\ 
 5 &  0.40 & 0.40 & 0.60 & 0.55 & 0.0 & 0.06 & 1.35 & 1.21 & 0.16 & 0.24 & 0.20 & 0.26 & 0.033 & 0.042 \\
 6 &  0.70 & 0.60 & 0.80 & 0.70 & 0.02 & 0.01 & 2.70 & 2.60 & 7.40 & 9.97 & 0.23 & 0.28 & 0.037 & 0.046\\ 
 7 &  0.43 & 0.38 & 1.2 & 1.5 & 0.01 & 0.07 & 3.50 & 4.02 & 1.39 & 1.89 & 0.24 & 0.30 & 0.038 & 0.049\\
 \hline
 & & & & & & & & & & & & Total & 0.09 & 0.10
\\
\hline
\label{table:oi-iso-uncerts}
\end{tabular}
\end{table*}

Similar results for the [O~{\sc iii}] optical and IR line fluxes falling within the PACS-IFU apertures are presented in Table~\ref{table:oiii-pacs}. The PACS spectra do not cover the 52 $\mu$m line, so the 52/88 $\mu$m density diagnostic is not available. To evaluate predicted 5007 \AA /88 $\mu$m line flux ratios for each IFU aperture, we adopted electron densities corresponding to those found for the nearest, usually overlapping, LWS apertures. So for
PACS-IFU apertures 1, 3, 4, 5, 6, 7, 8 and 9 we adopted the electron densities listed in Table~\ref{table:oiii52-iso} for LWS apertures 1, 7, 6, 6, 5, 2, 3 and 3, respectively.  PACS-IFU aperture 7 overlaps LWS aperture 2 in the northern region of Cas~A, for which we have argued that all of the interstellar dust column is in front of the remnant. The ratio of its ISM-dereddened 5007 \AA\ flux listed in column 5 of Table~\ref{table:oiii-pacs} to its 88 $\mu$m flux listed in column 2
of $8.96^{+2.10}_{-1.15}$ implies an electron temperature T$_e$ = 8100$^{+600}_{-400}$~K for n$_e$ = $1380^{+435}_{-420}$~cm$^{-3}$, in good agreement with the value of T$_e$ = 7900$^{+400}_{-700}$~K found for the overlapping LWS aperture 2. We adopted an electron temperature T$_e$ of 7900~K for all apertures in order to estimate the expected 5007 \AA\ line flux, F$_{5007,exp}$, using the 88 $\mu$m line flux.  
For regions of Cas~A other than 
PACS aperture 7 in the north we again assumed that half of the interstellar dust column lies in front of the remnant in order to calculate the ISM-dereddened 5007 \AA\ flux, F$_{5007,d}$.
The ratio of F$_{5007,exp}$ to F$_{5007,d}$ then yields the internal dust optical depths, $\tau_{int}$, that are listed in column 7. Table~\ref{table:oiii-pacs-uncerts} lists the statistical uncertainties on the measured and derived quantities in Table~\ref{table:oiii-pacs}.

The optical [O~{\sc iii}] spectra falling within apertures 1 and 9 of the PACS-IFU dataset show no obvious [O~{\sc iii}] 5007 \AA\ emission features. 
To calculate upper limits for the 5007 \AA\ fluxes in these apertures,
we used the largest peak in the aperture 6 spectrum, with a flux of 1.4$\times10^{-13}$~ergs~$cm^{-2} s^{-1}$, as a test emission feature. We found that co-adding a spectrum containing only this peak feature, scaled by factors of 0.2 and 0.3 to the spectra in apertures 1 and 9, respectively, resulted in spectra where the peak of the feature could just about be detected reliably. The resulting fluxes are adopted as 3-$\sigma$ upper limits in Table~\ref{table:oiii-pacs}.

We also have available for comparison optical and infrared line fluxes from a second ion, namely the 6300 \AA\ and 63 $\mu$m transitions of [O~{\sc i}].
The measured line fluxes are listed in Table~\ref{table:oi-iso}. Fluxes for the
63 $\mu$m line were measured only for the LWS apertures, since the PACS-IFU apertures did not provide enough detections of this line. The LWS observations did not provide enough detections of the [O~{\sc i}] 146 $\mu$m line for it to be usefully compared to the 63 $\mu$m line. The LWS observations of Cas~A included a position \citep[aperture 4; see Fig.~4 of][]{Docenko2010} which was offset to the northeast from the remnant. No [O~{\sc iii}] 52- or 88 $\mu$m emission was detected at this position but [O~{\sc i}] 63 $\mu$m emission was detected there, which we attribute to a diffuse interstellar source. We therefore subtracted the aperture~4 LWS spectrum from those obtained from the on-source apertures prior to measuring the net 63 $\mu$m line emission in the latter. Since the diffuse interstellar [O~{\sc i}] contribution may vary across Cas~A, this must be considered as an additional source of uncertainty - one which does not affect the [O~{\sc iii}] line flux measurements.

We can use the [O~{\sc i}] 63 $\mu$m line fluxes to predict the unreddened flux in the 6300 \AA\ line. This can then be compared to the ISM-dereddened 6300 \AA\ flux in order to evaluate the internal dust extinction in each aperture. The 6300 $\AA$/63 $\mu$m
flux ratio is insensitive to electron density but is sensitive to the electron temperature. For aperture 2, where all the projected interstellar dust column is expected to be in front of the remnant, the value of 0.85 for the ratio of the ISM-dereddened 6300 \AA\ flux to the observed 63 $\mu$m flux (Table ~\ref{table:oi-iso}) implies T$_e$ = 4800~K, significantly lower than the 7900-8100~K values derived from the
[O~{\sc iii}] optical to infrared line flux ratios. This discrepancy is probably because the ionized regions responsible for the 6300 \AA\ emission are not entirely responsible for the 63 $\mu$m flux, some of which may arise from cooler neutral regions within the remnant that do not excite [O~{\sc i}] 6300 \AA\ emission significantly. We have adopted the LWS aperture~2 6300 \AA /63 $\mu$m flux ratio of 0.85 in order to predict the unreddened 6300 \AA\ fluxes for the remaining LWS apertures. However, this is equivalent to assuming that all LWS apertures sample the same internal ionized to neutral gas fractions, and that the diffuse interstellar [O~{\sc i}] contribution does not vary from its offset aperture~4 level. 
The dust internal optical depths derived in Table~\ref{table:oi-iso} from the [O~{\sc i}] lines must therefore be considered more uncertain than those derived  in Tables~\ref{table:oiii52-iso} and 
\ref{table:oiii-pacs} from the [O~{\sc iii}] lines. Table~\ref{table:oi-iso-uncerts} lists the statistical uncertainties on the measured and derived quantities in Table~\ref{table:oi-iso}.

\subsection{The dust masses in the {\em ISO}-LWS and {\em Herschel} PACS-IFU projected apertures}

We will assume a simple case where no light has been scattered into the beams by internal dust, in which case the internal dust optical depths in Cas~A, deduced above for each aperture and listed in the penultimate columns of
Tables~\ref{table:oiii52-iso}, ~\ref{table:oiii-pacs} and ~\ref{table:oi-iso}, are given by
\begin{equation} \label{eq:dusteq1}
\tau_\lambda = \kappa_\lambda \times \rho \times l
\end{equation}
where $\kappa_\lambda$ is the dust opacity per unit mass at the wavelength observed, $\rho$ is the dust mass density and $l$ is the path length projected through Cas~A by each aperture. If $A$ is the area projected onto Cas~A by each aperture and 
$V$ = $A \times l$ is the volume projected through Cas~A by each aperture then we have
\begin{equation} \label{eq:dusteq2}
(\tau_\lambda/\kappa_\lambda) \times A = \rho \times V = M_d
\end{equation}

\begin{figure*}
  \includegraphics[width=.48\textwidth]{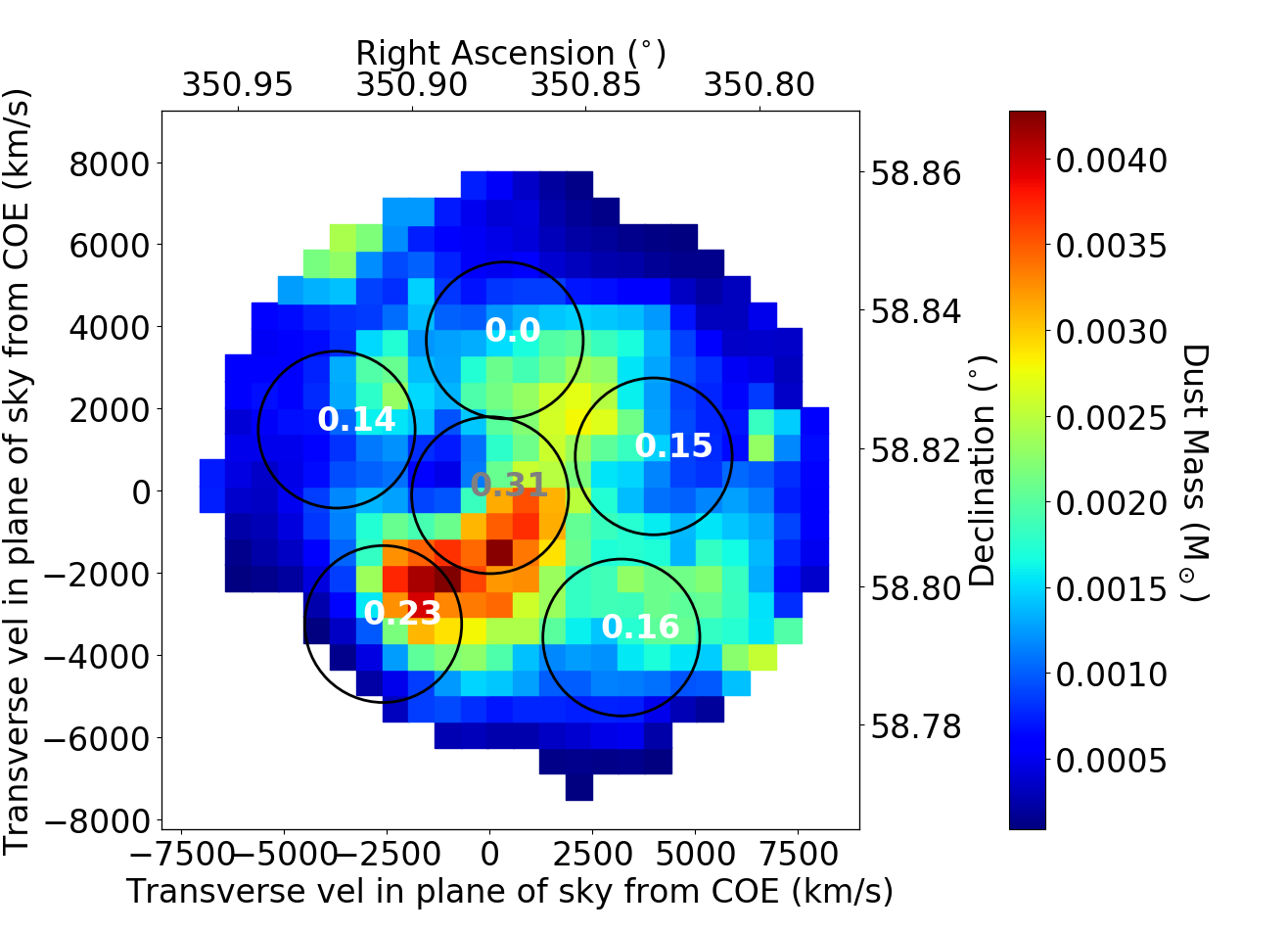} \qquad
  \includegraphics[width=.48\textwidth]{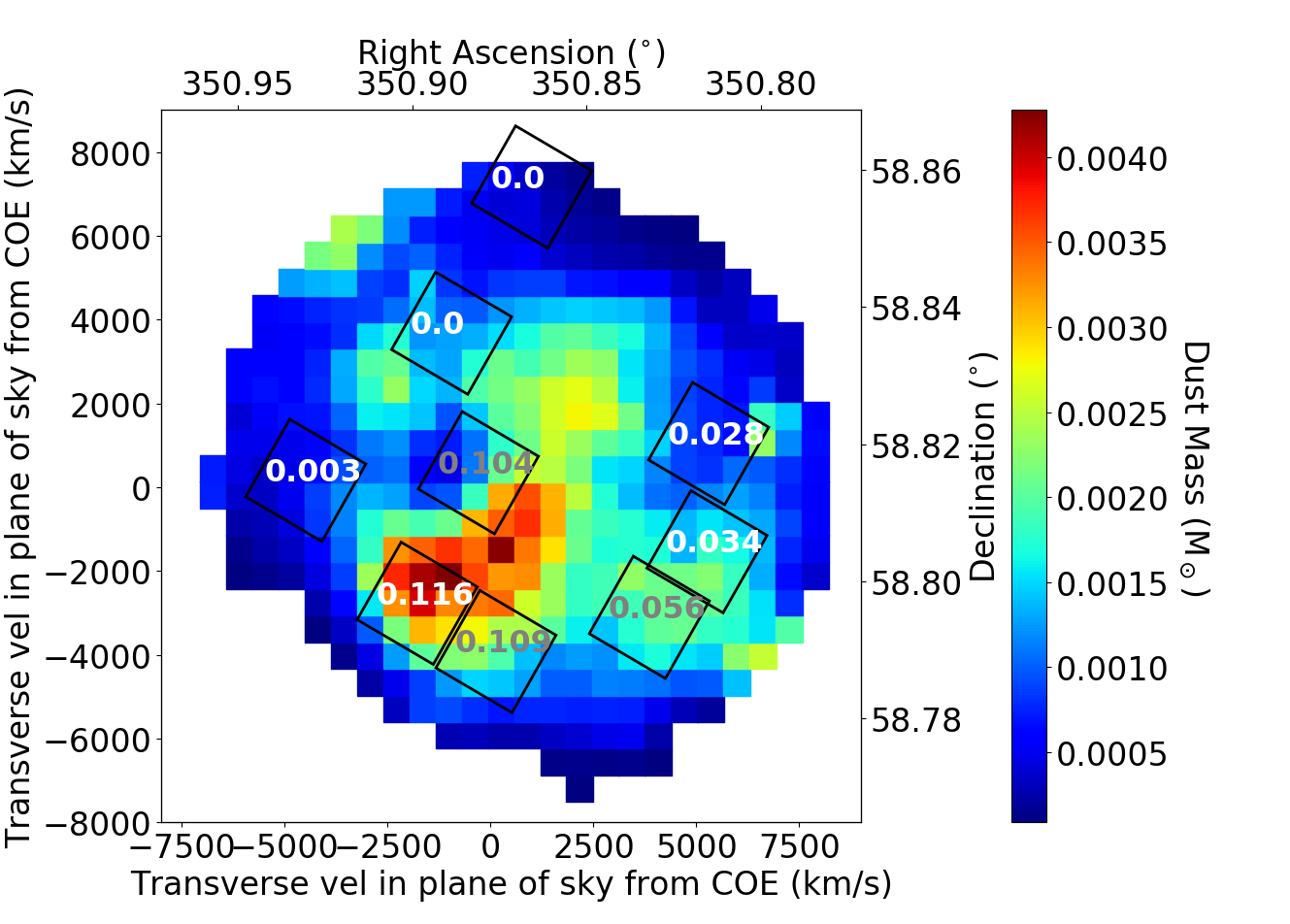}
  \caption{{\bf (A)} (left): Cas~A internal dust masses deduced from a comparison of [O~{\sc iii}] infrared and optical line fluxes falling within {\em ISO}-LWS apertures, labelled for each aperture with the dust mass, in solar masses, listed in Table~\ref{table:oiii52-iso}. They are overplotted on the resolved internal dust mass map of Cas A from Figure F3 of DL2017.  
{\bf (B)} (right): Internal dust masses deduced from a comparison of [O~{\sc iii}] infrared and optical line fluxes falling within PACS-IFU apertures, labelled for each aperture with the dust mass, in solar masses, listed in Table~\ref{table:oiii-pacs}. ``COE'' is an abbreviation for ``Centre Of Expansion''.}
 \label{fig:dustisopacsap}
\end{figure*}

\noindent
where $M_d$ is the mass of dust in the volume projected through Cas~A by each aperture. For consistency, the results in Tables \ref{table:oiii52-iso}, \ref{table:oiii-pacs} and \ref{table:oi-iso} use the same grain composition adopted by \cite{Bevan2017}: 50~per~cent silicate and 50~per~cent carbon grains with a grain size of 0.05 $\mu$m. However, in line with the work of \cite{Arendt_2014}, who found that silicates of a structure Mg$_{x}$SiO$_{x+2}$ best fitted the 21 $\mu$m feature of the {\em Spitzer} SED, we use Mg$_{2}$SiO$_{4}$ optical constants instead of the astronomical silicate constants used by \cite{Bevan2017}, although doing so only changes the derived dust mass by 1.4~per~cent, for the same 50:50 grain mixture as above. The effect on the derived dust mass of adopting different grain compositions is discussed further in Section~4.4 below.

As discussed above, we consider the internal dust mass estimates derived from the [O~{\sc iii}] line comparisons to be more reliable than those based on the [O~{\sc i}] line fluxes, so we will consider only our results from the former. The ratio of the total area subtended on Cas~A by the eight 47$\times$47-arcsec$^2$ PACS IFU aperture positions to that subtended by the six 84-arscec diameter circular LWS apertures is 0.53, while a value of $>$0.45 is obtained for the ratio our PACS-IFU summed dust mass of $>$0.45~M$_\odot$ to our LWS summed dust mass of 0.99~M$_\odot$.

The spatial distribution of internal dust from both the ISO-LWS and PACS apertures can be seen in Figure~\ref{fig:dustisopacsap}, overplotted on Figure F3 of DL2017, which shows the spatially resolved dust masses derived from the far-infrared and submillimetre {\em Herschel} images of the cold dust emission from the remnant. Our dust distribution closely follows the distribution found by DL2017: 
the largest internal dust columns are found towards the centre and southeast of Cas~A.
The LWS apertures did not cover all of Cas~A but in the dust mass map of \cite{DeLooze2017} 73~percent of the total dust mass was encompassed by the six LWS apertures.

For comparison with our LWS summed dust mass of 0.99~M$_\odot$,
\cite{DeLooze2017} derived a total mass of 0.5$\pm$0.1~M$_\odot$ for a 50:50 mixture of silicate and amorphous carbon dust grains within Cas~A while, from an analysis of the red-blue emission line asymmetries in the integrated optical spectrum of Cas~A, \cite{Bevan2017} derived a total
dust mass of $\sim$1.1~M$_\odot$ using the same 50:50 silicate:carbon dust mixture as adopted here.

\begin{figure*}
  \includegraphics[width=.48\textwidth]{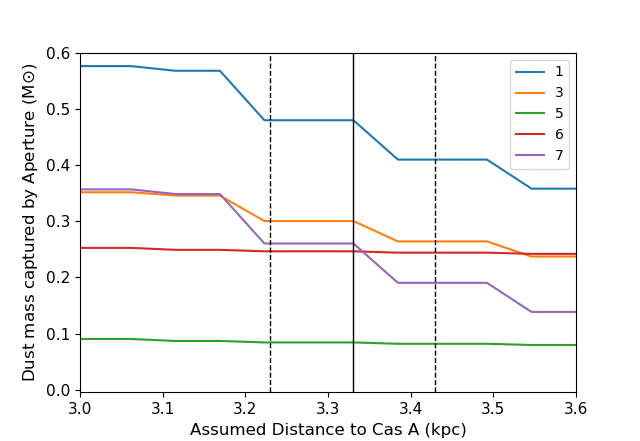} \qquad
  \includegraphics[width=.45\textwidth]{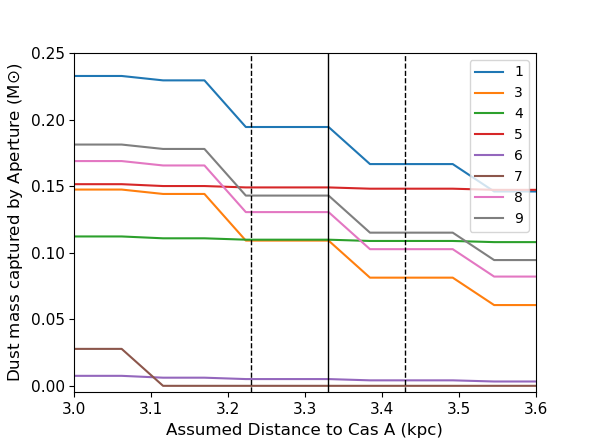}
  \caption{Dust masses within Cas~A based on the flux ratios of [O~{\sc iii}] optical to infrared lines emitted within projected {\em ISO}-LWS (left) and {\em Herschel}-PACS (right) apertures, as a function of position along the G2019 reddening versus distance curves for their lines of sight to Cas~A. The numbering of the curves refers to the numbering of the LWS and PACS-IFU apertures shown in Figures~\ref{fig:2d_casa_iso_op} and \ref{fig:2d_casa_pacs_op}. The black vertical line represents the distance of 3.33~kpc for Cas~A as found by \cite{Alarie2014}, and the black dashed lines indicate the stated uncertainty on their result of $\pm$0.10~kpc.}
  \label{fig:iso-pacs-dist}
\end{figure*}

\begin{figure*}
\includegraphics[width=\columnwidth]{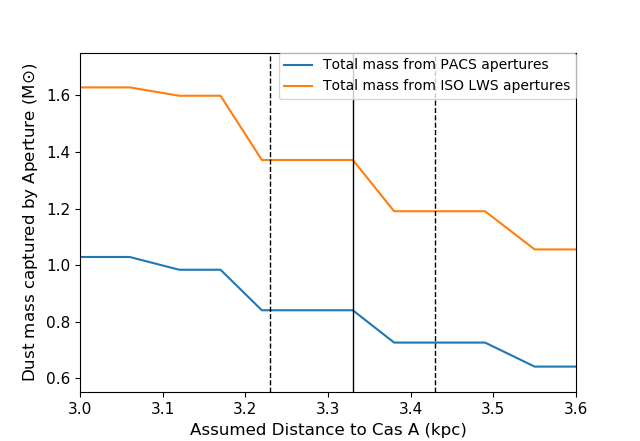}
\caption{Dust masses within Cas~A obtained by summing the masses contained within all projected {\em ISO}-LWS apertures (orange line) or {\em Herschel}-PACS apertures (blue line), as a function of position along the G2019 reddening versus distance curves for the lines of sight to Cas~A. The black vertical line represents the distance of 3.33~kpc for Cas~A as found by \cite{Alarie2014}, and the black dashed lines indicate the stated uncertainty on their result of $\pm$0.1~kpc.}
\label{fig:iso-pacs-tot}
\end{figure*}

\section{Dust mass sources of uncertainty}

\subsection{Measurement uncertainties}

Tables~\ref{table:oiii52-iso-uncerts}, ~\ref{table:oiii-pacs-uncerts} and \ref{table:oi-iso-uncerts} list the statistical uncertainties on the measured and derived quantities for the individual LWS and PACS apertures.
The uncertainties associated with the optical and IR line flux measurements were estimated from repeated measurements of the line fluxes above different choices for the background continuum levels. These measurement uncertainties were combined with the flux calibration uncertainties of 10~per~cent for the {\em ISO}-LWS spectra and 13~per~cent for the PACS-IFU spectra.
\\
\indent To estimate uncertainties associated with the foreground extinction measurements (listed in entries 6, 4 and 4 of Tables \ref{table:oiii52-iso-uncerts}, \ref{table:oiii-pacs-uncerts} and \ref{table:oi-iso-uncerts}, respectively), we calculated from the DL2017 interstellar extinction map A$_{\rm V}$ limits for the LWS and PACS apertures that corresponded to aperture radii that were 0.9 and 1.1 times their actual sizes. The uncertainties on the ISM-dereddened optical line fluxes, $F_{5007,d}$ and $F_{6300,d}$, were then obtained by combining the flux measurement uncertainties with the uncertainties associated with the foreground interstellar extinction estimates.
\\
The uncertainties on F$_{5007,exp}$ for the different {\em ISO}-LWS apertures are listed in Table~\ref{table:oiii52-iso-uncerts} and are a combination of the F$_{52}$ flux measurement uncertainties and the uncertainties associated with the adopted dust-free F$_{5007}$/F$_{52}$ ratio, derived in turn from the uncertainties on $n_{e}$ in each aperture and the uncertainty on the temperature of 7900~K calculated for LWS aperture 2.  
The F$_{5007,exp}$ uncertainties listed in Table \ref{table:oiii-pacs-uncerts} for the PACS apertures are a combination of the uncertainty on the dust-free F$_{5007}$/F$_{52}$ ratio derived for the {\em ISO}-LWS aperture~2, the F$_{88}$ flux measurement uncertainty for the PACS aperture, and the uncertainty associated with the F$_{52}$/F$_{88}$ ratio measured for the matching LWS aperture.
In Table~\ref{table:oi-iso-uncerts}, for the {\em ISO}-LWS apertures, the listed uncertainty for each F$_{6300,exp}$ value is a combination of the uncertainties associated with F$_{6300,d}$ and F$_{63}$ for LWS aperture~2, and the F$_{63}$ measurement uncertainty for the given aperture.
\\
\indent The uncertainties on $F_{\lambda,exp}$ and $F_{\lambda,d}$ enter weakly into the error budgets for the derived internal optical depths, which depend only logarithmically on line flux ratios. The scaling factor from a measured far-IR line flux to a predicted optical line flux depends exponentially on the adopted gas temperature but the logarithmic dependence of the dust internal depths on flux ratios means that the derived dust optical depths effectively have only a linear dependence on the adopted gas temperature. 
These weak dependencies may help explain why the LWS [O~{\sc i}]-based dust mass estimate of 0.80~M$_\odot$ (Table~\ref{table:oi-iso}) is relatively close to the LWS [O~{\sc iii}]-based dust mass estimate of 0.99~M$_\odot$ (Table~\ref{table:oiii52-iso}) despite the [O~{\sc i}] IR flux measurements and optical flux predictions being affected by greater uncertainties than those for the [O~{\sc iii}] lines, as discussed in Section~3.2. 

\subsection{Uncertainties associated with the location of the interstellar dust}

\begin{table*}
\caption{Dust absorption optical depths, $\tau_{abs}$, and total (absorption+scattering) dust optical depths, $\tau_{tot}$,  calculated for a projection of an {\em ISO}-LWS aperture through the centre of the \cite{Bevan2017}  best-fitting spherically symmetric model for Cas~A, compared with the effective optical depth, $\tau_{eff}$, deduced from the ratio of the output dust-affected to the input dust-free [O~{\sc iii}] $\lambda\lambda4959,5007$ \AA\ fluxes.}
\begin{tabular}{llllllc}
Clumped? & Dust mass (M$_\odot$) & \% silicate & $\tau_{abs}$ & $\tau_{tot}$ & $\tau_{eff}$ & $\tau_{eff}$/$\tau_{tot}$ \\

\hline
yes      & 1.1       & 50          & 0.38     & 0.46     & 0.49     & 1.06              \\
no       & 1.1       & 50          & 0.37     & 0.46     & 0.51     & 1.11              \\
no       & 5.5       & 50          & 1.88     & 2.31     & 2.04     & 0.88              \\
yes      & 5.5       & 50          & 1.86     & 2.28     & 1.44     & 0.63              \\
no       & 1.1       & 75          & 0.18     & 0.23     & 0.24     & 1.03              \\
yes      & 1.1       & 75          & 0.17     & 0.22     & 0.20     & 0.92              \\
no       & 5.5       & 75          & 0.87     & 1.14     & 1.17     & 1.02              \\
yes      & 5.5       & 75          & 0.86     & 1.12     & 0.89     & 0.80             
\end{tabular}
\label{table:bevan-taus}
\end{table*}

The source of uncertainty that we will discuss here is that associated with the correction of the optical line fluxes for foreground interstellar extinction. The factors by which the optical line fluxes have been corrected are large, ranging from 25 to 200 depending on position (see column~6 of Table~\ref{table:oiii52-iso} and column~4 of Table~\ref{table:oiii-pacs}), although this is then moderated by the logarithmic factor determining the internal optical depth estimate. For the above ISM extinction corrections, we have made use of the DL2017 ISM dust maps, based on their {\em Herschel} mapping of the far-IR emission by ISM dust in the vicinity of Cas~A. An alternative to their use that we discussed is to use the ISM extinction vs. distance curves available from the mapping of G2019, see Figs.~\ref{fig:bayestar-regions} and \ref{fig:extinct-sightlines}. 

Although the distance to Cas~A is
well established, the steepest parts of the G2019 extinction vs. distance curves in Fig~\ref{fig:extinct-sightlines} lie close to the position of Cas~A, so possible uncertainties in the exact range of distances over which the steep extinction rise take place can potentially have a significant effect on the adopted interstellar extinction corrections. To show the effect of shifting the extinction curves in Fig.~\ref{fig:extinct-sightlines} relative to the position of Cas~A, we illustrate in Fig.~\ref{fig:iso-pacs-dist} how the dust masses derived in each LWS and PACS-IFU aperture vary as the G2019 extinction curves are shifted by $\pm$0.3~kpc around the \cite{Alarie2014} distance to Cas~A of 3.33$\pm$0.10~kpc. We matched the G2019 extinction curve in the pink voxel in Figure \ref{fig:bayestar-regions} with ISO LWS aperture 2 and PACS aperture 7, the green voxel to ISO LWS aperture 1 and 7 and PACS apertures 1,8 and 9, the yellow voxel to ISO aperture 5 and 6, and PACS apertures 4,5 and 6. The extinction in ISO LWS aperture 3 was an average of the extinction in the green and yellow voxels. 

The total dust mass variation with adopted distance of Cas~A is seen in Figure~\ref{fig:iso-pacs-tot}.

Comparison of the DL2017-based total dust masses in Tables~\ref{table:oiii52-iso} and \ref{table:oiii-pacs} with the G2019-based total dust masses at 3.33~kpc shown in Figure~\ref{fig:iso-pacs-tot} shows the latter to be larger by factors of 1.5-2.0, which can be attributed to the generally lower ISM dust extinction corrections to the optical line fluxes yielded by the G2019 distance-extinction curves.

Figure~\ref{fig:extinct-sightlines} shows that for three of the four G2019 Cas~A sightlines the extinction starts to plateau at about 3~kpc, with the fourth plateauing just before 4~kpc.
The interstellar extinction in the G2019 map is deduced from stellar reddenings but, as discussed in Section~3.1, as the extinction increases with distance there comes a point beyond which there can be an inadequate number of detectable stars 
from which accurate distances and extinctions can be inferred, so that higher column densities of interstellar dust cannot be probed by the G2019 extinction curves. 
The G2019 extinction vs. distance curves are therefore likely to be plateauing at too low a distance due to their inability to probe higher dust extinctions.
As well as having much higher angular resolution, the DL2017 interstellar dust map, inferred from the optically thin far-infrared emission by the dust, can probe higher dust column densities. This explains why 
the deduced dust masses in the ISO-LWS and PACS apertures are lower when using the higher A$_{\rm V}$ values from the DL2017 interstellar extinction map to correct the optical fluxes for interstellar extinction. For the above reasons we consider our DL2017-based dust masses (Section 3.3) to be more reliable. 

We assumed earlier that a fraction $f = 0.5$ of the DL2017 interstellar extinction lies in front of the remnant for all of the projected LWS aperture areas apart from LWS-02, for which we adopted $f= 1.0$. For the five other LWS apertures, their derived enclosed total dust mass M$_d$ is found to vary with the value adopted for $f$ according to M$_d$ = (3.11 - 4.25$f$) M$_\odot$, yielding M$_d = 0.99~M_\odot$ for $f = 0.5$ and with M$_d$ falling to zero when $f$ reaches 0.73.

\begin{figure*}
  \includegraphics[width=.48\textwidth]{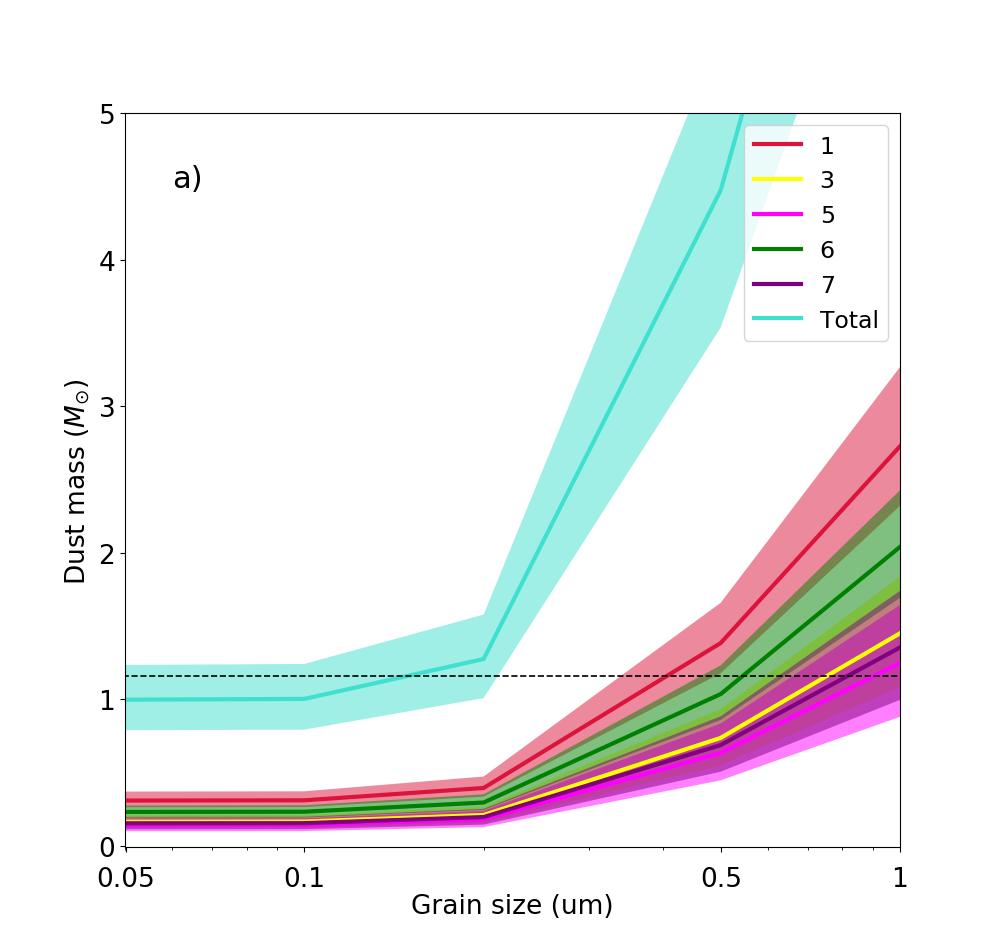} \qquad
  \includegraphics[width=.48\textwidth]{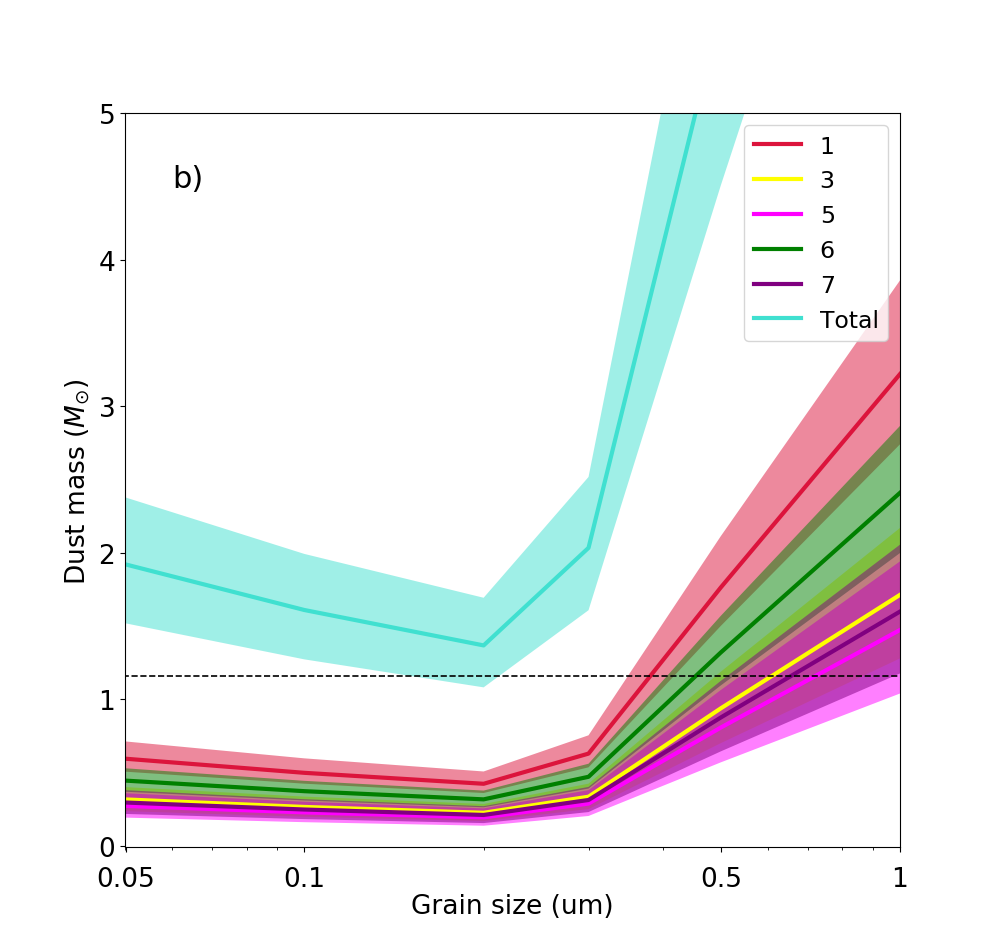}
\caption{a) The dust masses within the numbered {\em ISO}-LWS apertures as a function of grain radius for a 50:50 Mg$_2$SiO$_4$:Amorphous Carbon mix. 
b) Dust masses as a function of grain radius for a 75:25 Mg$_2$SiO$_4$:Amorphous Carbon mix.
The black dashed line corresponds to the maximum dust mass within Cas~A allowed by the nucleosynthetic yields of \cite{woosley2007}. }
\label{fig:silrat-mg2sio4}
\end{figure*}

\subsection{Uncertainties associated with the dust screening assumption}

Another source of uncertainty is that we have effectively assumed a foreground screen model for the dust, rather than the dust being mixed with the emitting gas, and light scattered back into the line of sight by dust has been neglected. 
\cite{Natta1984} have treated the case of multiple dust clumps, each with an optical depth $\tau_{cl}$, immersed in an emitting gas and find that for individual clump optical depths $<1$ 
then effective dust optical depths should be close to the actual optical depths.
The apparent overall optical depths at 5007 \AA\ measured for our LWS and PACS-IFU apertures range between 0.6 and 2.8 and individual dust clumps are expected to have low optical depths - if one adopts a clump radius of 10$^{16}$~cm \citep{Fesen2011} and a clump gas density of 100 oxygen atoms cm$^{-3}$ \citep{Kirchschlager2019}, together with a gas-to-dust mass ratio of 5-10 \citep{Priestley2019} along with a grain mass density of 3.2~g~cm$^{-3}$ and the grain radius of 0.05~$\mu$m from \cite{Bevan2017} used here, then one finds an optical depth from the centre to the surface of a clump of 0.125 - 0.25.

We have tested these expectations by running a number of {\sc damocles} gas plus dust Monte Carlo radiative transfer models for Cas~A, with parameters as described by \cite{Bevan2017}. The [O~{\sc iii}] emission models had an outer remnant radius R$_{out}$ of 5.2$\times10^{18}$~cm and an outer expansion velocity of 5000~km~s$^{-1}$, together with ratios of inner to outer radius and inner to outer velocity of 0.5. The dust and gas were mixed together and had $r^{-2}$ density distributions between the inner and outer radii. Models were run with smooth density distributions or clumped density distributions and with dust masses of either 1.1 or 5.5 M$_\odot$. For clumped models, a clump volume filling factor of f = 0.1 was used  with clumps of radius R$_{clump}$ = R$_{out}/25$.
Column~5 of Table~\ref{table:bevan-taus} lists each model's total (absorption+scattering) optical depth, $\tau_{tot}$ along a column projected through the centre of the remnant (corresponding to LWS aperture~1). Column~6 then lists $\tau_{eff}$, obtained via the ratio of the [O~{\sc iii}] 4959,5007 \AA\ dust-free input line flux to the dust-affected output line flux. Column~7 lists the ratio of $\tau_{eff}$/$\tau_{tot}$. For the four clumped dust models $\tau_{eff}$/$\tau_{tot} = 0.85\pm0.16$ while for the four models with smooth distributions of dust and gas we obtain $\tau_{eff}$/$\tau_{tot} = 1.01\pm0.08$. We have therefore made no corrections to the effective dust optical depths that have been derived in Tables~\ref{table:oiii52-iso}, \ref{table:oiii-pacs} and \ref{table:oi-iso}.

\subsection{Constraining the dust parameters}

\indent Cassiopeia A has been extensively studied at mid-IR wavelengths with {\em Spitzer}, which has enabled the silicate dust  species in Cas A to be well constrained \citep[]{Ennis_2006,Rho2008,Arendt_2014}.
We have determined the dust mass in the {\em ISO}-LWS apertures for a range of grain radii, silicate to amorphous carbon ratios and silicate dust species, in order to try to constrain the dust parameters in Cas A. \cite{Arendt_2014} found that Mg$_x$SiO$_{x+2}$ silicates provided the best fit to the 21 $\mu$m feature in the {\em Spitzer} SED. \cite{Rho2008} and \cite{douvion_2001} modelled the {\em Spitzer} spectra and {\em ISO}-SWS spectra respectively with MgSiO$_3$ as the predominent silicate composition. \cite{Arendt_2014} and \cite{Rho2008} also found that the addition of amorphous carbon to the silicate dust component improved their fits to the overall IR SED of Cas A. We calculated the dust masses in the {\em ISO} LWS apertures using Mg$_2$SiO$_4$, Mg$_{0.7}$SiO$_{2.7}$, Mg$_{2.4}$SiO$_{4.4}$ and MgSiO$_3$ as the silicate dust species, for silicate:amorphous carbon ratios of 0.5:0.5 and 0.75:0.25, with a range of grain radii from 0.05--1.0 $\mu$m. We used the silicate species optical constants of \cite{jager2003}, and the BE amorphous carbon optical constants of \cite{zubko1996}. 
\\
\indent Although different silicate species can have widely different
opacities at infrared wavelengths,
there is much less variation at visible wavelengths.
We found that the choice of silicate dust species did not greatly affect the derived dust masses. For the above silicate:carbon ratios, then for a given grain size the total derived dust mass in Cas~A varied by less than 15~per~cent between different Mg$_x$SiO$_{x+2}$ species, with a mean difference of 4~per~cent. 
The maximum difference between the dust masses derived using different silicate species was 60~per~cent, that between Mg$_2$SiO$_{4}$ and MgSiO$_{3}$ when using a grain radius of 0.3 $\mu$m. For silicate:carbon ratios of 0.5:0.5 and 0.75:0.25, the mean dust mass difference, over all grain radii from 0.05 to 1.0~$\mu$m, was 17~per~cent. For a fixed grain radius of 0.05 $\mu$m and a silicate:carbon ratio of 0.5:0.5, as per the results in Tables \ref{table:oiii52-iso}, \ref{table:oiii-pacs} and \ref{table:oi-iso}, we found that the total silicate+carbon dust masses varied by $<1$~per~cent when using the range of silicate species discussed above. As we do not consider these differences to be significant, for simplicity we show in Figure~\ref{fig:silrat-mg2sio4} only the dust masses that result from adopting Mg$_2$SiO$_4$ as the silicate dust species. 
\\
\indent Figure~\ref{fig:silrat-mg2sio4} illustrates how the total
silicate+carbon dust mass varies as the adopted grain radius is varied, for (a) a 50:50 ratio and (b) a 75:25 ratio of silicate to amorphous carbon.
The black dashed line plotted at 1.16~M$_\odot$ in both panels of Figure \ref{fig:silrat-mg2sio4} is the maximum dust mass of Mg$_2$SiO$_4$ and amorphous carbon in Cas~A predicted using the nucleosynthetic yields of \cite{woosley2007} for a 25~M$_\odot$ progenitor star. In panel b) of Figure \ref{fig:silrat-mg2sio4}, for the 75/25 Mg$_2$SiO$_4$ to amorphous carbon ratio, the only consistent grain radius is approximately 0.2 $\mu$m. This would give a dust mass of 1.36~M$_\odot$, where the maximum dust mass permissible within Cas~A given the nucleosynthetic limit is within the uncertainty limits. For the 50/50 ratio of panel a), all dust grain radii smaller than 0.2 $\mu$m are comfortably within the nucleosynthetic yield limit. 
\\
\indent Grain radii of 1 $\mu$m, which DL2017 adopted for their analysis of the far-infrared dust emission, would give a very large total dust mass here ($>5$~M$_\odot$), as such large grains are inefficient absorbers in the optical regime. However, DL2017 noted that the assumed size of the grains does not strongly affect the dust emissivity at IR/submm wavelengths and that adopting a smaller grain radius would only change their derived dust mass of 0.5$\pm$0.1~M$_\odot$ within the model uncertainties. Therefore the DL2017 results are consistent with those obtained here.
\section{The survivability of Cas~A's dust}
\indent An interesting question is the future survivability of the current mass of dust against destruction by shocks in Cas~A.
\indent \cite{Kirchschlager2020} treated sputtering, grain-grain collisions, ion trapping and gas accretion in Cas~A conditions and found that for an initial log-normal grain size distribution peaking at 0.1 $\mu$m, between 15 and 50~per~cent of silicate grains would survive an encounter with the reverse shock, for a range of clump density contrasts, while \cite{Kirchschlager2019} found that between 10-30~per~cent of amorphous carbon grains could survive. So for an initial 0.99~M$_\odot$ of a 50:50 silicate:carbon dust mix, 0.07--0.25~M$_\odot$ of silicates and 0.05--0.15~M$_\odot$ of carbon dust could survive to reach the ISM.
When considering dust destruction by sputtering only, \cite{slav2020} found that for initial grain radii of 0.1 $\mu$m or 0.2 $\mu$m, then respectively 3 or 10~per~cent of the mass of silicate dust and 30 or 45~per~cent of the mass of amorphous carbon dust would survive passage through the remnant shocks. So for an initial 50:50 silicate:carbon dust mass of 0.99~M$_\odot$ consisting of 0.1 $\mu$m radius grains, their results predict that 0.015~M$_\odot$ of silicates and 0.15~M$_\odot$ of amorphous carbon grains would survive, while for a 75:25 silicate:carbon total dust mass of 
1.36~M$_{\odot}$ consisting of 0.2 $\mu$m radius grains, their results predict that 0.10~M$_{\odot}$ of Mg$_2$SiO$_4$ and 0.15~M$_{\odot}$ of amorphous carbon would survive shock impacts. \cite{Morgan2003} and \cite{Dwek2007} have estimated that a typical CCSN would need to produce $\geq$0.1~M$_\odot$ of dust to account for a significant fraction of cosmic dust, a constraint that the above estimates satisfy, giving further weight to the hypothesis that CCSNe can be major contributors to galactic dust budgets.

\section{Conclusions}

{\em ISO}-LWS and {\em Herschel}-PACS
spectroscopic measurements of far-infrared [O~{\sc iii}] and [O~{\sc i}] emission line fluxes at multiple positions across Cas~A have been compared with measurements of the optical line fluxes from the same ions in equivalent apertures. Maps of interstellar dust extinction versus distance for different sightlines towards Cas~A were used to correct the measured optical line fluxes for foreground dust extinction. For the northern region of Cas~A the interstellar dust column was estimated to lie entirely in front of the remnant. After correcting for this foreground extinction, the 5007 \AA /52 $\mu$m line flux ratio yielded an electron temperature for the O$^{2+}$ emitting gas of T$_e$ = 7900$^{+400}_{-700}$~K.
This value of T$_e$ was adopted for the other observed positions, along with the electron densities implied by the measured [O~{\sc iii}] 52/88 $\mu$m ratios at each position, in order to predict the expected 5007 \AA\ flux at each position from the observed [O~{\sc iii}] far-IR line fluxes at the same positions. The ratio of these expected 5007 \AA\ fluxes to the observed values after foreground extinction corrections had been applied enabled internal dust optical depths to be evaluated at each aperture position on the remnant. These dust optical depths were converted to dust masses by using optical constants for a 50:50 silicate:amorphous carbon mixture of 0.05 $\mu$m radius grains. 

From a comparison of {\em ISO}-LWS [O~{\sc iii}] 52 $\mu$m fluxes with M2013 5007 \AA\ fluxes from the same aperture areas, a summed dust mass of 0.99$\pm$0.1~M$_\odot$ within the projected apertures was derived. 
The dust mass in Cas~A derived by the IR-optical emission-line method used here agrees within the quoted uncertainties with those from two other independent methods, namely the IR SED-based dust mass estimate of 0.5$\pm$0.1~M$_\odot$
by \cite{DeLooze2017} and the optical emission line red-blue asymmetry-based dust mass estimate of 1.1~M$_\odot$ by \cite{Bevan2017}.
\citet{Laming2020} have questioned whether the dust mass within Cas~A could be as large as 0.5-1.1~M$_\odot$, preferring the IR SED-based dust mass of $\leq$0.1~M$_\odot$ obtained by \citet{Arendt_2014}, but we note that the latter's SED analysis ignored the {\em Herschel}-SPIRE photometric data available for Cas~A longwards of 200~$\mu$m, a spectral region where its coolest and most massive dust component emits strongly.

Since high-resolution 3D optical data exist for the LMC remnant N132D \citep{Law2020} and for the SMC remnant 1E 0102.2-7219 \citep{Vogt2017}, future spaceborne infrared missions could provide resolved far-IR spectra of these and similar young remnants to which the method dscribed here could be applied. {\em Herschel} PACS-IFU and {\em ISO}-LWS spectra exist for the Crab Nebula, so an analysis using our method would also be possible given suitable resolved optical spectra.

\section*{Data Availability}

The infrared spectra used here are available from the ESA {\em ISO} and {\em Herschel} archives, at \url{http://archives.esac.esa.int/nida/} and \url{http://archives.esac.esa.int/hsa/whsa/} respectively. Copies of the Cas~A optical spectroscopic dataset can be obtained from D.M. upon reasonable request.

\section*{Acknowledgements}

MND, MJB and AB acknowledge support from European Research Council (ERC) Advanced Grant 694520 SNDUST. D.M. acknowledges National Science Foundation support from grants PHY-1914448 and AST-2037297. IDL acknowledges support from European Research Council (ERC) Starting Grant 851622 DustOrigin.




\bibliography{Casapaper} %


\label{lastpage}
\end{document}